\documentclass[11pt]{article}

\usepackage[final]{acl}

\usepackage{times}
\usepackage{latexsym}
\usepackage{amsmath}
\usepackage{amssymb}
\usepackage{booktabs}
\usepackage{tabularx}
\usepackage{multirow}
\usepackage{colortbl}
\usepackage{wrapfig}

\usepackage[T1]{fontenc}

\usepackage[utf8]{inputenc}

\usepackage{microtype}

\usepackage{inconsolata}

\usepackage{graphicx}
\usepackage{xspace}
\usepackage{subcaption}
\usepackage{fvextra}
\usepackage[most]{tcolorbox}

\definecolor{promptframe}{HTML}{8B4513}
\definecolor{promptbody}{HTML}{FBEEE1}

\newtcolorbox{promptbox}[1]{
  enhanced, breakable,
  colback=promptbody, colframe=promptframe,
  colbacktitle=promptframe, coltitle=white,
  fonttitle=\bfseries, title=#1,
  arc=2mm, outer arc=2mm,
  boxrule=0.5pt, titlerule=0pt,
  left=6pt, right=6pt, top=4pt, bottom=4pt,
  boxsep=2pt,
}

\newcommand{\Librarian}{Librarian\xspace}

\newif\ifrevtrack
\revtrackfalse
\definecolor{revcolor}{HTML}{C2185B}
\newcommand{\rev}[1]{\ifrevtrack{\color{revcolor}#1}\else{#1}\fi}
\newcommand{\revon}{\ifrevtrack\color{revcolor}\fi}

\title{Long Live the Librarian! A Persistent Search Sub-Agent \\ for Energy-Efficient Multi-Agent Software Engineering Systems}

\author{
 \textbf{Seunghyuk Cho\textsuperscript{1}},
 \textbf{Sunghyun Choi\textsuperscript{1}},
  \textbf{Jaeseung Heo\textsuperscript{1}},
 \textbf{Youngbin Choi\textsuperscript{1}}, 
 \\
 \textbf{Saemi Moon\textsuperscript{2}},
 \textbf{MoonJeong Park\textsuperscript{1}},
 \textbf{Dongwoo Kim\textsuperscript{1,2,*}}
\\
 \textsuperscript{1}Graduate School of Artificial Intelligence, POSTECH,
\\
 \textsuperscript{2}Department of Computer Science and Engineering, POSTECH, 
\\
 \small{
    \{shhj1998, sunghyunchoi, jsheo12304, choi.youngbin, saemi, mjeongp, dongwoo.kim\}@postech.ac.kr
 }
}

\begin{document}
\maketitle
\begin{abstract}
Multi-agent systems (MAS) have substantially advanced autonomous software engineering (SWE), but their growing inference energy demands raise sustainability concerns.
In this paper, we demonstrate that this cost is concentrated in an overlooked source: redundant output tokens generated across agents.
Two empirical findings ground this claim.
First, our per-token energy attribution for MAS reveals a sharp asymmetry: an output token consumes 30 to 1,000 times more energy than an input or cached token.
Second, MAS inflate per-episode output because agents repeatedly re-explore overlapping repository regions.
To address this inefficiency, we propose \Librarian, a persistent search sub-agent that tracks repository-search history and suppresses redundant exploration actions across agents. 
By returning short references to file regions instead of full file excerpts, \Librarian further reduces output-token volume. 
On SWE-Bench Verified \rev{and SWE-Bench-Live}, \Librarian reduces per-episode GPU energy consumption of existing multi-agent SWE systems by \rev{11 to 30\%} while preserving task performance.
Our code is available at \url{https://github.com/ml-postech/Librarian}.
\end{abstract}

\section{Introduction}

The rapid deployment of large language models (LLMs) has driven a sharp rise in the electricity demand of large-scale computing and is projected to drive further growth in the coming years~\citep{energyfootprint1, energyfootprint2, fernandez-etal-2025-energy, morrison2025holistically}. This consumption translates into a substantial environmental footprint, ranging from carbon emissions~\citep{luccioni2024} to freshwater drawn by host data centers~\citep{li2025}.
These concerns have motivated recent work on measuring LLM inference energy~\citep{fernandez-etal-2025-energy, niu2026tokenpowerbench} and improving efficiency through system- and hardware-aware optimizations~\citep{patel2025, stojkovic2025dynamollm}.

However, most existing analyses focus on the per-call setting and offer limited insight into agentic deployments where a single task spans many LLM calls. This limitation is sharpest in multi-agent systems (MAS) for software engineering (SWE)~\citep{tao2024magis, hong2024metagpt, wadhwa2024masai, qian-etal-2024-chatdev, phan2024hyperagent, xu2026boad}, where bug fixing~\citep{jimenez2024swebench, zhang2026swebench} and feature implementation~\citep{hong2024metagpt, qian-etal-2024-chatdev} are split across specialized sub-agents. 
MAS are increasingly adopted in SWE for the performance gains of role specialization, yet no energy analysis exists for this setting. SWEnergy~\citep{tripathy2025swenergyempiricalstudyenergy} takes a first step but is restricted to single-agent runs, leaving MAS deployments uncharacterized.

In this work, we measure the GPU energy of multi-agent SWE systems and find a counterintuitive pattern: MAS consume more GPU energy than their single-agent counterparts even at lower aggregate token counts. To explain this gap, we conduct turn-level attribution that decomposes each episode's GPU energy into the input, output, and cached tokens of every LLM call. The attribution traces the gap to two compounding effects: MAS produce more output tokens per episode, and each output token incurs much higher energy than its input or cached counterpart. Thus, aggregate token counts and API-pricing-based proxies can underestimate actual GPU energy and obscure the category-level effects that drive agentic energy consumption.

We trace the excessive output to redundant file exploration across sub-agent invocations. In existing MAS designs, no sub-agent preserves an exploration history across invocations, so each lookup begins from scratch. As a result, when the orchestrator issues different queries, sub-agents repeatedly traverse overlapping files and regenerate long file excerpts, amplifying the most energy-intensive token category.

To break this re-exploration cycle, we propose the \Librarian, a MAS-agnostic code-lookup sub-agent that returns the relevant region for each orchestrator query. We instantiate it with three design choices. First, to eliminate redundant exploration, its session persists across all invocations within an episode, so prior searches are reused. Second, to keep the session consistent with the current repository state, each invocation is supplied with the files and line ranges edited since the previous call, prompting the \Librarian to re-traverse those regions. Third, to minimize output tokens, each answer is returned as a short view command rather than the inlined file excerpt itself. 
To integrate \Librarian into an existing MAS, we provide a guideline that replaces the MAS's code-navigation sub-agent with \Librarian and routes all exploration through it.

We apply the \Librarian on SWE-Bench Verified~\citep{jimenez2024swebench} \rev{and SWE-Bench-Live~\citep{zhang2026swebench}} to BOAD~\citep{xu2026boad} and HyperAgent~\citep{phan2024hyperagent} with two open-weight LLMs from the Qwen3.6 family, 35B-A3B~\citep{qwen36_35b_a3b} and 27B~\citep{qwen36_27b}. Across both MAS and both backbones, the \Librarian variants preserve or improve the underlying MAS's pass rate while reducing the GPU energy consumed per episode by \rev{11 to 30\%}. Ablations confirm that this energy reduction stems from \Librarian invocations eliminating the duplicate output tokens diagnosed above.

We summarize the contributions as follows: we confirm that output tokens dominate resource use in agentic settings (\S\ref{sec:energy-characterization}); we diagnose the energy overhead of existing MAS designs as a consequence of duplicate lookups across the debugging cycle that no existing sub-agent recognizes (\S\ref{sec:redundancy-source}); we propose the \Librarian, a persistent search sub-agent that integrates into existing MAS designs to supply this missing role (\S\ref{sec:method}); and we show on SWE-Bench Verified \rev{and SWE-Bench-Live} that it \rev{reduces} hardware energy at matched pass rate (\S\ref{sec:experiments}).

\begin{table}[t!]
\centering
\setlength{\tabcolsep}{5pt}
\renewcommand{\arraystretch}{1.15}
\resizebox{\linewidth}{!}{
\begin{tabular}{@{}lrrrrr@{}}
\toprule
& \multicolumn{4}{c}{Token Count} & \\
\cmidrule(lr){2-5}
Method & Uncached & Cached & Out & Total & Energy \\
& (M) & (M) & (K) & (M) & (kJ) \\
\midrule
\rowcolor{gray!12}[0pt][0pt]
\multicolumn{6}{@{}l@{}}{\textbf{Qwen3.6-35B-A3B}} \\
Vanilla & 0.11 & 2.69 & 16.8 & 2.81 & 10.31 \\
HyperAgent & 0.26 & 2.16 & 48.5 & 2.47 & 25.64 \\
BOAD & 0.19 & 3.71 & 28.5 & 3.93 & 17.45 \\
\midrule
\rowcolor{gray!12}[0pt][0pt]
\multicolumn{6}{@{}l@{}}{\textbf{Qwen3.6-27B}} \\
Vanilla & 0.10 & 2.59 & 15.4 & 2.71 & 57.45 \\
HyperAgent & 0.21 & 1.35 & 35.4 & 1.59 & 121.84 \\
BOAD & 0.21 & 3.07 & 24.3 & 3.30 & 88.96 \\
\bottomrule
\end{tabular}
}

\caption{Token usage and energy consumption across SWE Agents. We report token usage by category, uncached input, cached input, and output, alongside per-episode mean energy consumption for a single-agent SWE system denoted as Vanilla and two multi-agent SWE systems, HyperAgent and BOAD. HyperAgent shows the smallest cached and total token counts yet the largest energy consumption.}
\label{tab:total-vs-energy}
\vskip -0.1in
\end{table}

\section{Related Work}

\subsection{Energy footprint of LLM inference}

LLM inference energy is conventionally quantified at the GPU level, since GPU consumption dominates per-server serving power~\citep{patel2024characterizing}.
Within this scope, \citet{luccioni2024} benchmark per-inference energy across model and task configurations, and \citet{fernandez-etal-2025-energy} further model per-query GPU duration and energy as a function of inference-optimization choices and workload geometry.
At a finer granularity, \citet{niu2026tokenpowerbench} decompose per-request power into the prefill and decode stages, yielding per-token energy estimates per stage that inform energy-aware serving designs~\citep{patel2025, stojkovic2025dynamollm}.
While these works yield hardware-grounded per-call energy characterizations, all are defined per inference call, not over multi-turn agentic trajectories.
SWEnergy~\citep{tripathy2025swenergyempiricalstudyenergy} extends per-call energy measurement to SWE-agent frameworks, but its analysis is restricted to small LLMs with near-zero task resolution.
Moreover, its per-category attribution uses univariate correlation, which gives each category's per-token energy scaled by how much its count varies across turns, not the per-token energy itself.

\subsection{Multi-agent software engineering systems}

Multi-agent SWE systems organize execution around specialized roles such as planner, navigator, coder, and reviewer that collaborate to resolve a single task. Two paradigms for stitching these roles recur. Pipeline-based MAS hardcode the role sequence at design time, so each role advances the task one predetermined phase and hands off to the next role~\citep{qian-etal-2024-chatdev, hong2024metagpt, wadhwa2024masai}. In SWE, however, execution order is not fixed in advance and coordination evolves as the task unfolds~\citep{xu2026boad}. Orchestrator-driven MAS instead route through a central orchestrator that picks the next sub-agent at each step, so the workflow emerges at runtime~\citep{tao2024magis, phan2024hyperagent, xu2026boad}. Such designs form sub-agent teams at runtime to fit each issue~\citep{tao2024magis}.

\subsection{Context management of LLMs}

Context management for LLMs extracts query-relevant content from model history or external sources. Retrieval-based methods rank documents or prior turns by similarity to the current query and prepend the top results to one LLM call~\citep{lewis2020rag, zhang-etal-2023-repocoder, shrivastava2023repofusiontrainingcodemodels}. Compression-based methods remove or summarize non-essential tokens from an oversized prompt to keep query-aligned content~\citep{jiang2023llmlingua, jiang-etal-2024-longllmlingua, chevalier2023adapting, li-etal-2023-compressing}. Agentic memory persists state across many calls of a single agent through structured memory stores or reflective summaries~\citep{packer2024memgptllmsoperatingsystems, shinn2023reflexion, park2023generative, xu2026amem}. Their target, however, is a single agent interacting with a user or environment, rather than an orchestrator-invoked sub-agent within MAS.

\section{An Energy Profile of SWE Agents}
\label{sec:energy-characterization}

In this section, we identify where the GPU energy of SWE agents is spent.
We first describe how we measure per-turn token counts and GPU energy.
We then decompose per-turn energy into the contributions of three
token categories and show that output tokens dominate. Finally, we
trace the excess output of MAS to repeated file
lookups across sub-agent invocations.

\subsection{Measuring per-turn energy and tokens}
\label{sec:preliminaries}

We treat each SWE-agent turn as a sequence of an LLM call, action extraction, and action execution. A SWE-agent's episode is a
sequence of turns, and per-episode cost sums over turns. This applies
to single-agent systems and MAS alike; the two architectures differ
only in how each turn's context is assembled.

For each turn $t$ we record two quantities. The first is a triple of
token counts: the input tokens processed by prefill, $x^u_t$; the
input tokens served from the key-value cache, $x^c_t$; and the output
tokens generated by decoding, $x^o_t$. The second is the GPU energy
drawn during the call, $E_t$, obtained by reading NVIDIA's hardware
energy counter and subtracting the
idle baseline.\footnote{NVML exposes this counter on the A100-80GB
and B200 SXM GPUs used in this study.} Appendix~\ref{sec:per-turn-statistics}
states the measurement formally.

\subsection{Token-level energy attribution}
\label{sec:token-level-energy-attribution}

We first ask whether aggregate token count predicts energy.
Table~\ref{tab:total-vs-energy} shows that it does not: one of the systems achieves the smallest total token count yet draws the most energy. We
therefore need a finer-grained attribution.

To isolate each token category's contribution, we fit a multiple
linear regression of per-turn energy on the three token counts:
\begin{equation}
\label{eq:energy-model}
E_t = \alpha + \beta_u x^u_t + \beta_c x^c_t + \beta_o x^o_t + \epsilon_t,
\end{equation}
where $\alpha$ absorbs fixed per-turn overhead, $\beta_i$ is the
marginal energy per token of category $i$, and $\epsilon_t$ is the
residual. We fit this regression on per-turn records collected from
Vanilla, BOAD~\citep{xu2026boad}, and HyperAgent~\citep{phan2024hyperagent}
runs on SWE-Bench Verified~\citep{jimenez2024swebench}, across two
backbones, Qwen3.6-35B-A3B~\citep{qwen36_35b_a3b} and
Qwen3.6-27B~\citep{qwen36_27b}, and two GPUs, A100 and B200. Full
fit details appear in
Appendix~\ref{sec:linear-regression-energy}.

Table~\ref{tab:per-token-energy} reports the fitted coefficients, and Table~\ref{tab:energy-diag} reports the variance-inflation factors (VIFs), which diagnose coefficient reliability. Output tokens cost 30 to 1{,}000 times more energy per token than either input category, and all VIFs satisfy $\mathrm{VIF} \le 1.05$, indicating nearly orthogonal regressors and thus reliable coefficient estimates. The same ordering appears at the episode level: in Table~\ref{tab:total-vs-energy}, per-episode energy across systems tracks output token counts rather than total or input counts. Output volume is therefore the dominant lever for reducing per-episode energy.

\begin{table}[t!]
\centering
\setlength{\tabcolsep}{5pt}
\renewcommand{\arraystretch}{1.15}
\resizebox{\columnwidth}{!}{%
\begin{tabular}{@{}r@{ / }l r@{ $\pm$ }l r@{ $\pm$ }l r@{ $\pm$ }l c@{}}
\toprule
\multicolumn{2}{c}{} & \multicolumn{6}{c}{Per-token energy (mJ/tok)} & \\
\cmidrule(lr){3-8}
\multicolumn{2}{c}{Setting} & \multicolumn{2}{c}{$\beta_u$} & \multicolumn{2}{c}{$\beta_c$} & \multicolumn{2}{c}{$\beta_o$} & $R^2$ \\
\midrule
A3B & A100 & $30.50$ & $1.95$ & $1.36$ & $0.11$ & $\mathbf{967}$ & $\mathbf{20}$ & $0.979$ \\
A3B & B200 & $6.43$ & $1.56$ & $0.54$ & $0.02$ & $\mathbf{660}$ & $\phantom{0}\mathbf{8}$ & $0.987$ \\
27B & B200 & $33.97$ & $1.04$ & $2.04$ & $0.06$ & $\mathbf{3,512}$ & $\mathbf{23}$ & $0.992$ \\
\bottomrule
\end{tabular}%
}

\caption{Estimated per-token energy by token category. 
We fit a linear regression of total per-request energy on the number of uncached input ($u$), cached input ($c$), and output tokens ($o$); $\beta_u$, $\beta_c$, $\beta_o$ are the resulting per-token coefficients and $R^2$ is the regression fit. 
Output tokens
show a 30 to 1{,}000$\times$ gap relative to the other categories.}
\label{tab:per-token-energy}
\end{table}

\subsection{Sources of excess MAS output}
\label{sec:redundancy-source}

\S\ref{sec:token-level-energy-attribution} shows that BOAD and
HyperAgent emit more output per episode than the
single-agent Vanilla. To localize this excess, we examine how
per-turn output relates to the agent's file-access pattern. A large
share of MAS output is emitted on turns whose file read repeats a
region the sub-agent has already inspected earlier in the episode.

The mechanism is straightforward. The file read itself is an
external observation, not output. The output arrives when the
sub-agent then re-emits the located region, quoted or paraphrased,
in the answer it returns to the orchestrator. A region inspected
multiple times across the episode is therefore rewritten multiple
times.

\paragraph{\rev{Bounding duplicate output.}}
We flag a turn whose action is a file read as a duplicate read if the returned observation overlaps, in file-line range, an earlier read within the same episode; any intervening write to the file resets this history. \rev{The output tokens emitted on such turns give an \emph{upper bound on duplicate output}, the output that re-states content the sub-agent has already read, since a flagged turn may also emit new reasoning that our accounting cannot separate out. We report this bound rather than a point estimate, and we discuss the tightness of the gap in Appendix~\ref{app:dup-bound}.} We further split duplicate reads into same-agent overlap, where the prior read came from the same sub-agent invocation, and cross-agent overlap, where it came from a different invocation. Two invocations of the same sub-agent role count as cross-agent, because read history is scoped to an invocation rather than to a role.

Table~\ref{tab:dup-breakdown} \rev{breaks this bound down by the origin of the overlapping earlier read, each cell as a percentage of the system's file-read output tokens,} for BOAD and HyperAgent, aggregated across both backbones.
Both systems show frequent within-invocation re-exploration, and HyperAgent additionally shows across-invocation re-exploration even within the same role. Together, these indicate that re-exploration accounts for a non-trivial portion of the MAS baselines' excess output.

\begin{table}[t]
  \centering
  \resizebox{\linewidth}{!}{
  \begin{tabular}{llrrrr}
    \toprule
    & & \multicolumn{3}{c}{Across invocations (\%)} & \multirow{3}{*}[-3pt]{\shortstack{Within\\invocation\\(\%)}} \\
    & & \multicolumn{3}{c}{Source} & \\
    \cmidrule(lr){3-5}
    System & Current & \multicolumn{1}{c}{$R_1$} & \multicolumn{1}{c}{$R_2$} & \multicolumn{1}{c}{$R_3$} & \\
    \midrule
    \multirow{3}{*}{BOAD}
      & $R_1$ & 0.0 & 8.1 & 0.1 & 9.1 \\
      & $R_2$ & 0.1 & 0.0 & 0.1 & 18.6 \\
      & $R_3$ & 9.9 & 3.7 & 0.0 & 22.7 \\
    \midrule
    \multirow{3}{*}{HyperAgent}
      & $R_1$ & 12.1 & 0.3 & 0.9 & 9.8 \\
      & $R_2$ & 1.9 & 0.4 & 0.3 & 0.7 \\
      & $R_3$ & 1.8 & 0.3 & 26.8 & 34.4 \\
    \bottomrule
  \end{tabular}
  }
  \caption{\rev{Upper bound on duplicate output in BOAD and HyperAgent. Each cell shows the fraction of the file-read output tokens that are duplicated, measured as an upper bound. \emph{Current} is the role performing the repeated read, and the columns under \emph{Source} give the origin of the earlier overlapping read. $R_1$, $R_2$, and $R_3$ are the code navigator, issue analyzer, and orchestrator for BOAD, and the navigator, editor, and executor for HyperAgent. \emph{Within invocation} counts an earlier read made in the same invocation of the current role, whereas \emph{Across invocations} counts an earlier completed invocation of the source role.}}
  \label{tab:dup-breakdown}
  \vskip -0.1in
\end{table}



\section{Method: The Librarian Sub-Agent}
\label{sec:method}

\subsection{Overview}
\label{sec:method:overview}

We propose the \Librarian, a persistent sub-agent that handles
every code lookup an episode requires through a single long-lived
session, eliminating the within-episode re-exploration diagnosed in
\S\ref{sec:redundancy-source}. \rev{Four} properties define it.
\textbf{Persistent session:} one session spans the whole episode, so
every query starts from the full history of prior searches.
\textbf{Pointer-only answers:} each answer to the orchestrator is a
list of view-command pointers rather than inlined file content,
keeping output tokens small in the response. \textbf{\rev{Lookup}-only scope:} the role
is restricted to file and code lookup, excluding the analytical and
editing work that MAS navigators typically conflate with
it~\citep{xu2026boad}. \rev{\textbf{Freshness report:} each invocation is
supplied with the files and line ranges edited since the previous call,
keeping the persisted history consistent with the current repository
state.}

\S\ref{sec:method:invocation} walks through a single invocation and
shows how each property takes effect. \S\ref{sec:method:adapt} provides
a MAS-agnostic recipe for integration. Figure~\ref{fig:librarian_architecture}
visualizes the architecture, and the implementation details appear in
Appendix~\ref{app:librarian}.

\begin{figure*}[t!]
\centering
\includegraphics[width=\linewidth]{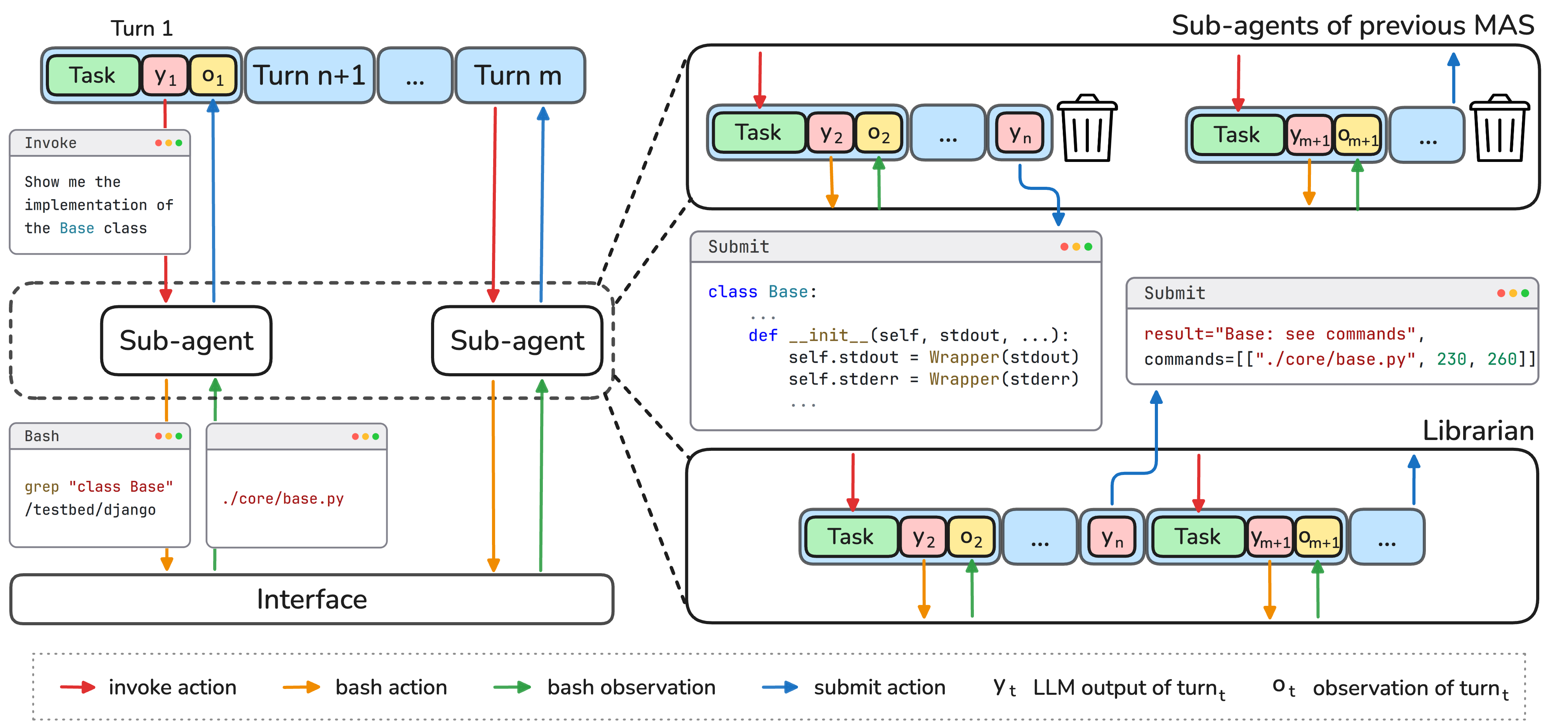}
\caption{Illustration of the overall process of multi-agent SWE systems resolving a SWE task.
The orchestrator calls a sub-agent via \texttt{invoke} action, which navigates the repository through \texttt{bash} and file view/edit tools and returns its answer via a \texttt{submit} call; the content of this call becomes the observation for the orchestrator's invocation.
On a subsequent invocation of the same role, prior MAS (top right) start from a fresh context, discarding all history, and returns the full file excerpt.
In contrast, the \Librarian (bottom right) appends the new task given by the invocation to its existing trajectory and continues from there, reusing the accumulated repository knowledge, and only generates the file pointer for response, which is converted to file excerpt later.
}
\label{fig:librarian_architecture}
\vskip -0.05in
\end{figure*}

\subsection{Anatomy of an invocation}
\label{sec:method:invocation}

An invocation proceeds in three stages \rev{that together realize the
properties} defined in \S\ref{sec:method:overview}.

\paragraph{Context assembly.}
The \Librarian's context retains its full trajectory from earlier
invocations in the episode, so prior searches and answers are on
hand. To keep this history aligned with the repository's current
state, each invocation prepends a \emph{freshness report} listing
the files and line ranges modified since the previous call, and
instructs the \Librarian to re-traverse those regions before reusing
cached excerpts. \rev{Invocations that surface no previously unseen file
content} are dropped from the persisted history. The implementation details of pruning the invocation history and freshness report appear in Appendix~\ref{app:librarian:pruning} and Appendix~\ref{app:librarian:freshness}.
For context management, we also try retrieval instead of history accumulation, comparing sparse and dense top-$k$ variants over the same prior history in \S\ref{sec:exp-retrieval}.

\paragraph{Scoped search.}
The \Librarian then resolves the query by running search commands
such as \texttt{grep} and \texttt{find} and by viewing matching
files, under two scope guards. The system prompt declares editing,
test execution, and analytical reasoning out of scope\rev{, and routes
every file read through OpenHands's file view
tool~\citep{wang2025openhands} so that each read is recorded at an
explicit file-line range. The tool layer backs this up by stripping
that tool and the \texttt{bash} shell of every mutating operation}.

\paragraph{Pointer-only submission.}
The invocation closes with a call to a \texttt{submit} tool. Its
argument is a list of view commands referring to the located code,
plus a natural-language explanation; the commands cannot inline the
code itself. The tool executes the commands on the \Librarian's
behalf and forwards the excerpts to the orchestrator. The
orchestrator therefore receives the actual snippets, while the
\Librarian's output stays a short pointer list.
An example of a submission and the orchestrator message it produces appears in Appendix~\ref{app:librarian:submission}.

\subsection{Integration into an existing MAS}
\label{sec:method:adapt}

An existing orchestrator's system prompt typically comprises three
blocks: a SWE context describing the operating environment, the
per-task user query, and a \emph{solve plan} laying out the
step-by-step procedure. Adding the \Librarian touches only the
solve plan; the SWE context, the user query, and every
non-navigation step of the plan stay untouched.

\paragraph{Tool swap and plan rewrite.}
We replace the orchestrator's existing navigator-delegation tool
with a \Librarian-delegation tool, then rewrite each navigation step
of the solve plan to phrase its action as a natural-language lookup
query. Any navigator responsibility beyond file and code lookup,
such as cross-module analysis, is redirected to other steps in the
plan. The rewrite is formalized as a prompt-level guideline in
Appendix~\ref{app:librarian:rewrite-guide}, and the resulting BOAD
and HyperAgent plans appear in Appendix~\ref{app:librarian:plans}.

\paragraph{Tool-selection policy.}
We prepend a short policy to the solve plan to keep the \Librarian
invoked only for lookup. The policy directs the orchestrator to
call the \Librarian for unknown-location queries and to keep
analytical tasks such as bug diagnosis outside the \Librarian's
scope. \rev{It takes two forms, differing only in whether the
orchestrator can view files itself; both appear in}
Appendix~\ref{app:librarian:tool-policy}.

\section{Experiments}
\label{sec:experiments}

In this section, we empirically validate the \Librarian's ability to reduce per-episode energy consumption. We also confirm that the saving stems from lower output tokens, the primary mechanism that the \Librarian's design targets.

\subsection{Experimental setup}
\label{sec:exp-setup}

We describe the experimental setup used to compare the \Librarian against prior systems on the SWE task, covering the benchmark, implementation details, and baselines.

\subsubsection{Benchmark}
We evaluate every method on the 500 tasks of SWE-Bench Verified~\citep{jimenez2024swebench}, a benchmark of real GitHub issues drawn from open-source Python repositories. For each task, the SWE-agent receives the issue description and the repository at the buggy commit, and must produce a patch that passes the maintainer's test suite for the issue.
\rev{To test whether the findings hold outside the repositories and issues covered by SWE-Bench Verified, we additionally evaluate on 100 tasks sampled from the verified split of SWE-Bench-Live~\citep{zhang2026swebench}, which follows the same task format but is built from recent GitHub issues over a continually updated and considerably larger pool of repositories.}

\subsubsection{Implementation details}
\label{sec:exp-impl}

We explain the two main components to run a SWE-agent: i) the LLM driving it and ii) the scaffold mediating its repository interaction.

\paragraph{LLM serving.}
For the LLM, we evaluate two Qwen3.6 variants: the sparse mixture-of-experts 35B-A3B~\citep{qwen36_35b_a3b} with $3$B active parameters and the dense 27B~\citep{qwen36_27b}. Both backbones run in FP8 on a single NVIDIA B200 SXM GPU, served by vLLM~\citep{kwon2023efficient}. We enable prefix caching so that each sub-agent reuses its prior prefix across successive invocations within a session. We size the KV cache to avoid eviction in every episode. The serving configuration and the sampling hyper-parameters are held fixed across all episodes as listed in Appendix~\ref{app:hyperparameters}.

\paragraph{Agent interface.}
For the interface which allows the SWE-agent to interact with the code repository, we provide a base tool set comprising a \texttt{bash} shell, the \texttt{str\_replace\_editor} file editor from OpenHands~\citep{wang2025openhands} that takes a view or edit mode along with a file path and a line range, and a \texttt{submit} tool that marks the agent's solution as final. 
Each agent role draws on whichever subset of this set it needs and may additionally define role-specific tools such as sub-agent delegation. To keep long-trajectory runs stable, we adopt three safeguards from the OpenHands runtime, a wall-clock timeout on each shell command, loop detection on agent outputs, and truncation of tool observations. We describe how each mechanism is implemented in Appendix~\ref{app:hyperparameters}.

\subsubsection{Baselines}
\label{sec:exp-baselines}

We confirm the \Librarian's effectiveness by applying it to two orchestrator-delegation multi-agent SWE systems. We additionally compare against token-efficiency methods that can be combined with any SWE-agent.

\paragraph{Multi-agent systems.}
We consider two MAS designs that delegate the SWE workflow across specialized sub-agents: BOAD~\citep{xu2026boad} and HyperAgent~\citep{phan2024hyperagent}.
BOAD's orchestrator first delegates code localization for the issue to one sub-agent, then bug-cause navigation to a second sub-agent, and finally produces a patch by aggregating the outputs of both sub-agents.
HyperAgent's orchestrator manages three sub-agents: a navigator for code exploration, an editor for file modification, and an executor for test execution. The orchestrator itself holds only the sub-agent delegation and submit tools.
Following a pre-defined plan detailed in Appendix~\ref{app:librarian:plans}, the orchestrator iterates this navigate-edit-execute cycle until it submits a patch.
While BOAD imposes no tool restrictions on its sub-agents, HyperAgent restricts each sub-agent's tool set by role, such as confining the navigator to file viewing and the editor to file edits.

\paragraph{Token-efficiency methods.}
We consider two token-efficiency techniques: caveman prompting~\citep{brussee2026caveman, hakim2026brevityconstraintsreverseperformance} and LastNObservation~\citep{yang2024sweagent} that we apply across every role of the MAS to reduce token consumption.
Caveman prompting appends a fixed style directive, given in Appendix~\ref{app:caveman}, to the system prompts of every sub-agent and the orchestrator, inducing simpler answer text while leaving the thinking trace intact.
LastNObservation is a SWE-agent~\citep{yang2024sweagent} scaffold feature that truncates all but the most recent $N$ tool observations in each role's context, replacing each truncated observation with a short placeholder while preserving the thinking traces and action calls. We set $N=5$ in all subsequent experiments.

\subsection{Results and analysis}
\label{sec:exp-main}

We first report the main results of attaching the \Librarian to existing MAS baselines on SWE-Bench Verified, then analyze the observed energy reduction from \rev{four} angles: \rev{whether it holds on a wider set of repositories,} \rev{the trend of the energy reduction across task difficulty,} localization to the replaced role, and the reuse of accumulated history across invocations.

\paragraph{Results.} 
Table~\ref{tab:main-results} reports results on SWE-Bench Verified for BOAD and HyperAgent across four configurations: bare MAS, MAS with a token-efficiency technique, \Librarian-augmented MAS, and \Librarian-augmented MAS with caveman prompting.
We report pass rate, per-episode output tokens, and GPU energy, denoted as Pass, Out tok, and Energy respectively.
The \Librarian matches or improves the pass rate in most of the settings, while substantially reducing energy.
\rev{Paired tests over the 500 tasks confirm both effects with high confidence: the output-token and energy reductions are significant in 7 of the 8 comparisons, and the pass rate is preserved in all 8 within a 3.5-point margin. Appendix~\ref{app:significance} details the test designs and reports every comparison.}
LastNObservation truncates recent observations to reduce per-turn input tokens, but per-episode GPU energy increases across all settings: sub-agents are forced to re-explore lost context, inflating output tokens in three of four settings. Pass rate also degrades in every setting. Caveman prompting alone reduces energy but degrades pass rate in three of four settings.
Across all settings, the \Librarian with caveman prompting attains the lowest energy consumption.

\begin{table}[t!]
\centering
\resizebox{\linewidth}{!}{
\begin{tabular}{lrrr}
\toprule
Method & Pass & Out tok & Energy \\
 & (\%) & (K) & (kJ) \\
\midrule
\rowcolor{gray!12}
\multicolumn{4}{l}{\textbf{Qwen3.6-35B-A3B}} \\
HyperAgent & 66.0 & 48.5 & 25.64 \\
\hspace{1ex}$+$ Librarian & \textbf{67.2} & 42.2 & 22.73 \\
\hspace{1ex}$+$ LastNObservation & 56.8 & 53.3 & 34.79 \\
\hspace{1ex}$+$ Caveman & 66.2 & 37.7 & 19.99 \\
\hspace{1ex}$+$ Caveman $+$ Librarian & 66.2 & \textbf{34.7} & \textbf{18.63} \\
\midrule
BOAD & 69.8 & 28.5 & 17.45 \\
\hspace{1ex}$+$ Librarian & \textbf{71.6} & 24.6 & 14.66 \\
\hspace{1ex}$+$ LastNObservation & 63.2 & 33.5 & 26.87 \\
\hspace{1ex}$+$ Caveman & 68.4 & 24.7 & 14.79 \\
\hspace{1ex}$+$ Caveman $+$ Librarian & 70.4 & \textbf{20.6} & \textbf{12.37} \\
\midrule
\rowcolor{gray!12}
\multicolumn{4}{l}{\textbf{Qwen3.6-27B}} \\
HyperAgent & 69.8 & 35.4 & 121.84 \\
\hspace{1ex}$+$ Librarian & \textbf{72.0} & 26.3 & 91.45 \\
\hspace{1ex}$+$ LastNObservation & 64.8 & 34.0 & 135.84 \\
\hspace{1ex}$+$ Caveman & 68.0 & 21.9 & 79.40 \\
\hspace{1ex}$+$ Caveman $+$ Librarian & 67.6 & \textbf{18.0} & \textbf{66.05} \\
\midrule
BOAD & \textbf{71.8} & 24.3 & 88.96 \\
\hspace{1ex}$+$ Librarian & \textbf{71.8} & 19.7 & 71.76 \\
\hspace{1ex}$+$ LastNObservation & 70.2 & 26.2 & 122.91 \\
\hspace{1ex}$+$ Caveman & 70.2 & 17.5 & 66.20 \\
\hspace{1ex}$+$ Caveman $+$ Librarian & 71.4 & \textbf{13.5} & \textbf{51.21} \\
\bottomrule
\end{tabular}
}

\caption{Results on \textit{SWE-Bench Verified}. We report the overall pass rate alongside the per-episode means for output tokens and energy consumption for each SWE agent. The Librarian improves output token efficiency and energy consumption across all settings while preserving the pass rate.}
\label{tab:main-results}
\vskip -0.19in
\end{table}

\paragraph{\rev{Results across a wider set of repositories.}}
\rev{We run the same protocol on 100 tasks sampled from SWE-Bench-Live, which span 46 repositories, 41 of them absent from SWE-Bench Verified.
Table~\ref{tab:live-results} reports the result on Qwen3.6-27B: the \Librarian reduces output tokens and energy on both MAS at a comparable pass rate.}

\begin{table}[t!]
\centering
\revon
\setlength{\tabcolsep}{5pt}
\renewcommand{\arraystretch}{1.15}
\resizebox{\linewidth}{!}{
\begin{tabular}{lrrr}
\toprule
Method & Pass (\%) & Out tok (K) & Energy (kJ) \\
\midrule
\rowcolor{gray!12}
\multicolumn{4}{l}{\textbf{Qwen3.6-27B}} \\
HyperAgent & \textbf{23.0} & 49.5 & 182.32 \\
\hspace{1ex}$+$ Librarian & 21.0 & \textbf{34.9} & \textbf{128.26} \\
\midrule
BOAD & 23.0 & 29.9 & 116.88 \\
\hspace{1ex}$+$ Librarian & \textbf{25.0} & \textbf{21.5} & \textbf{82.57} \\
\bottomrule
\end{tabular}
}

\caption{\rev{Results on \textit{SWE-Bench-Live}. We report the per-episode means over 100 tasks sampled from the verified split, with Qwen3.6-27B as the backbone. The Librarian reduces output tokens and energy on both systems at a comparable pass rate.}}
\label{tab:live-results}
\end{table}

\begin{figure}[t!]
\centering
\includegraphics[width=.75\linewidth]{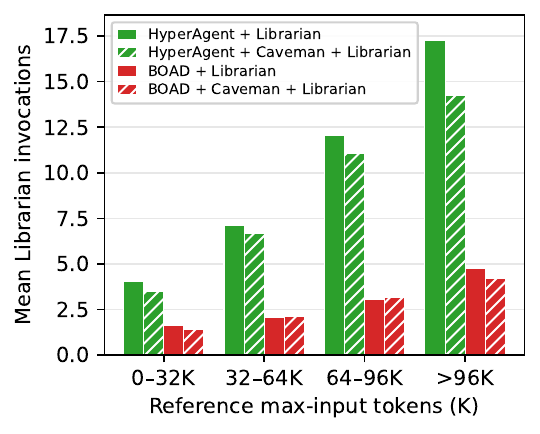}
\caption{Mean Librarian invocations per episode across task difficulty. We plot the per-episode mean Librarian invocation count for the Librarian-augmented MAS variants run with Qwen3.6-27B against the reference max-input-token bin. The invocation frequency grows monotonically with task difficulty. The plot for 35B-A3B is in Figure~\ref{fig:librarian-wakes-by-bin}.}
\label{fig:librarian-wakes-by-bin_main}
\vskip -0.1in
\end{figure}

\paragraph{\rev{Energy reduction across task difficulty.}}
\rev{We examine how the energy reduction varies with task difficulty.}
We define each task's difficulty as the max input tokens observed on a reference single-agent run.
Table~\ref{tab:difficulty-energy} stratifies per-episode energy by difficulty: \rev{the reduction is present in every bin} and is largest on the harder tasks, with the \Librarian variants attaining the lowest energy in \rev{14 of the 16 cells}.
\rev{The reduction} persists even as \rev{the \Librarian} is invoked more often on harder tasks, as shown in Figure~\ref{fig:librarian-wakes-by-bin_main}.

\begin{table}[t!]
\centering
\resizebox{\linewidth}{!}{
\begin{tabular}{lrrrr}
\toprule
\multicolumn{1}{c}{} & \multicolumn{4}{c}{Reference max-input tokens (K)} \\
\cmidrule(lr){2-5}
\multirow{2}{*}{Method} & 0--32K & 32--64K & 64--96K & >96K \\
 & (n=36) & (n=312) & (n=130) & (n=22) \\
\midrule
\rowcolor{gray!12}
\multicolumn{5}{l}{\textbf{Qwen3.6-35B-A3B}} \\
HyperAgent & 10.36 & 17.61 & 40.46 & 76.87 \\
\hspace{1ex}$+$ Librarian & 5.83 & 16.20 & 36.84 & 59.65 \\
\hspace{1ex}$+$ LastNObservation & 7.90 & 26.25 & 56.29 & 72.69 \\
\hspace{1ex}$+$ Caveman & \textbf{4.33} & 12.92 & 33.11 & 68.30 \\
\hspace{1ex}$+$ Caveman $+$ Librarian & 4.88 & \textbf{12.42} & \textbf{31.00} & \textbf{56.10} \\
\midrule
BOAD & 7.22 & 12.50 & 26.99 & 47.97 \\
\hspace{1ex}$+$ Librarian & 5.12 & 11.03 & 20.98 & 44.41 \\
\hspace{1ex}$+$ LastNObservation & 7.69 & 19.82 & 43.28 & 60.83 \\
\hspace{1ex}$+$ Caveman & 4.31 & 10.28 & 23.64 & 43.64 \\
\hspace{1ex}$+$ Caveman $+$ Librarian & \textbf{3.92} & \textbf{8.82} & \textbf{19.04} & \textbf{36.97} \\
\midrule
\rowcolor{gray!12}
\multicolumn{5}{l}{\textbf{Qwen3.6-27B}} \\
HyperAgent & 32.61 & 79.36 & 208.19 & 359.95 \\
\hspace{1ex}$+$ Librarian & 25.31 & 64.40 & 142.91 & 279.35 \\
\hspace{1ex}$+$ LastNObservation & 37.35 & 106.47 & 217.76 & \textbf{229.48} \\
\hspace{1ex}$+$ Caveman & 25.93 & 54.11 & 129.45 & 231.94 \\
\hspace{1ex}$+$ Caveman $+$ Librarian & \textbf{19.05} & \textbf{41.17} & \textbf{110.24} & 234.59 \\
\midrule
BOAD & 40.68 & 69.77 & 127.81 & 210.46 \\
\hspace{1ex}$+$ Librarian & 34.78 & 56.26 & 104.22 & 160.26 \\
\hspace{1ex}$+$ LastNObservation & 52.24 & 101.43 & 174.50 & 238.36 \\
\hspace{1ex}$+$ Caveman & 26.20 & 48.93 & 100.29 & 175.19 \\
\hspace{1ex}$+$ Caveman $+$ Librarian & \textbf{20.10} & \textbf{37.43} & \textbf{80.67} & \textbf{123.42} \\
\bottomrule
\end{tabular}
}

\caption{Per-episode GPU energy (kJ) of the SWE-agents with respect to the task difficulty. We report the per-episode GPU energy of the SWE-agents across difficulty classes. The addition of the Librarian reduces the energy consumption in almost all settings.}
\label{tab:difficulty-energy}
\end{table}

\paragraph{Energy reduction at the lookup role.}
We investigate the source of the energy reduction, focusing on the role that the \Librarian replaces.
For each MAS baseline, we identify the role the \Librarian substitutes for: BOAD's code navigator and HyperAgent's navigator.
Table~\ref{tab:role-comparison-compact} reports that the \Librarian outputs fewer tokens and consumes less energy than its counterpart.
Tables~\ref{tab:role-comparison-hyperagent} and~\ref{tab:role-comparison-boad} further show that this reduction does not inflate the token usage or energy consumption of other roles.

\begin{table}[t!]
\centering
\setlength{\tabcolsep}{6pt}
\renewcommand{\arraystretch}{1.15}
\resizebox{\linewidth}{!}{
\begin{tabular}{@{}llrrcrr@{}}
\toprule
 & & \multicolumn{2}{c}{Replaced role} & & \multicolumn{2}{c}{$+$ Librarian} \\
\cmidrule(lr){3-4} \cmidrule(lr){6-7}
Model & Method & Out (K) & E (kJ) & & Out (K) & E (kJ) \\
\midrule
\multirow{4}{*}{A3B} & HyperAgent & 13.92 & 7.54 & & \textbf{6.02} & \textbf{3.57} \\
 & HyperAgent $+$ Caveman & 10.51 & 5.69 & & \textbf{4.67} & \textbf{2.66} \\
 & BOAD & 13.81 & 8.54 & & \textbf{1.93} & \textbf{1.08} \\
 & BOAD $+$ Caveman & 11.08 & 6.62 & & \textbf{1.70} & \textbf{0.94} \\
\midrule
\multirow{4}{*}{27B} & HyperAgent & 10.50 & 36.63 & & \textbf{4.70} & \textbf{16.92} \\
 & HyperAgent $+$ Caveman & 6.99 & 25.74 & & \textbf{3.38} & \textbf{12.65} \\
 & BOAD & 5.63 & 20.79 & & \textbf{1.82} & \textbf{6.56} \\
 & BOAD $+$ Caveman & 5.10 & 19.32 & & \textbf{1.44} & \textbf{5.43} \\
\bottomrule
\end{tabular}
}

\caption{Per-episode comparison on the role that the Librarian replaces. We report the per-episode means of output tokens and GPU energy at the replaced role. The Librarian reduces energy at the replaced role across all settings. The full per-role breakdown is in Appendix~\ref{app:role-breakdown}.}
\label{tab:role-comparison-compact}
\end{table}

\begin{figure}[t!]
\centering
\input{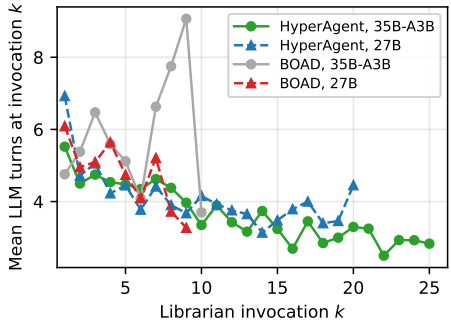}
\vspace{-10pt}
\caption{Visualization of the mean LLM turns to answer per \Librarian invocation. We report the mean number of turns each \Librarian variant takes to emit an output at the $k$-th invocation. All cases except BOAD on 35B-A3B show a decreasing trend as $k$ grows.}
\label{fig:librarian-turns-per-wake}
\vskip -0.1in
\end{figure}

\paragraph{History reuse across invocations.}
We track the number of LLM turns the \Librarian takes to answer the orchestrator's $k$-th call, averaged over episodes that reach the $k$-th call. Figure~\ref{fig:librarian-turns-per-wake} plots the average number of turns against $k$. The number of turns decreases in most configurations over $k$, confirming that the \Librarian reuses its accumulated history to resolve later queries with fewer turns.

\subsection{\rev{Ablation of the \Librarian's components}}
\label{sec:exp-ablation}

\rev{The \Librarian bundles the four mechanisms of \S\ref{sec:method:overview}: the persistent session, pointer-only answers, the lookup-only scope, and the freshness report.
We first quantify the contribution of each mechanism with a leave-one-out ablation, then examine which of them are specific to the navigation role.}

\paragraph{\rev{Leave-one-out ablation.}}
\rev{Each variant removes exactly one mechanism and leaves the rest unchanged, on both MAS with Qwen3.6-27B over 100 tasks sampled from SWE-Bench Verified. Appendix~\ref{app:librarian:noscope} specifies how the variants are built.
Table~\ref{tab:ablation-loo} reports the pass rate together with the \emph{increase} in per-episode output tokens and energy caused by the removal.
Removing any one of the four raises both output tokens and energy on both systems, so every mechanism contributes to the reduction.}

\begin{table}[t!]
\centering
\revon
\setlength{\tabcolsep}{4pt}
\renewcommand{\arraystretch}{1.15}
\resizebox{\linewidth}{!}{
\begin{tabular}{lrrrrrr}
\toprule
& \multicolumn{3}{c}{HyperAgent} & \multicolumn{3}{c}{BOAD} \\
\cmidrule(lr){2-4}\cmidrule(lr){5-7}
	Removed mechanism & Pass (\%) & $\Delta$out (K) & $\Delta E$ (kJ) & Pass (\%) & $\Delta$out (K) & $\Delta E$ (kJ) \\
\midrule
--- (full \Librarian) & 68.0 & --- & --- & 70.0 & --- & --- \\
\midrule
Persistent history & 71.0 & $+5.9$ & $+20.2$ & 70.0 & $+1.3$ & $+3.3$ \\
Pointer-only output & 71.0 & $+17.4$ & $+62.4$ & 72.0 & $+3.4$ & $+11.5$ \\
Lookup-only scope & 72.0 & $+10.5$ & $+37.7$ & 71.0 & $+3.9$ & $+13.9$ \\
Freshness report & 69.0 & $+6.7$ & $+23.5$ & 70.0 & $+1.6$ & $+5.1$ \\
\bottomrule
\end{tabular}
}

\caption{Leave-one-out ablation of the Librarian's four mechanisms on 100 SWE-Bench Verified tasks with Qwen3.6-27B. $\Delta$out and $\Delta E$ are the increases in per-episode output tokens and GPU energy caused by removing the mechanism. All the four mechanisms contribute on the reduction of output token and energy.}
\label{tab:ablation-loo}
\end{table}

\paragraph{\rev{Transfer to a non-navigation role.}}
\rev{Pointer-only answers, the lookup-only scope, and the freshness report are all defined over file lookups, whereas the persistent session is not tied to a particular role.
We therefore give the persistent session alone to HyperAgent's executor, which reproduces the issue and runs tests, on 100 SWE-Bench Verified tasks with Qwen3.6-27B and everything else unchanged.
Per-episode output tokens fall from $42.9$K to $34.3$K and energy from $147$ to $127$\,kJ, while the paired pass counts differ by 2 tasks out of 100.
The persistent session is therefore not specific to code navigation.}

\subsection{Comparison with retrieval baselines}
\label{sec:exp-retrieval}

We compare the \Librarian against two retrieval-based variants that replace the persistent history with retrieved context. The variants use BM25~\citep{robertson2019bm25} for sparse scoring and SFR-Embedding-Code-2B\_R~\citep{liu2024codexembed} for dense scoring.
At each invocation, the scoring method ranks prior file-read command-observation pairs by similarity to the orchestrator's query, and the top-$k$ form the context.
Table~\ref{tab:retrieval-comparison} reports pass rate, output tokens, and per-episode energy across the four variants on a shared random sample of 100 tasks per backbone with $k=5$.
\rev{The \Librarian attains the lowest energy with a pass rate at or above both retrieval baselines in 3 of the 4 settings}, indicating that the orchestrator's queries need active context management and reasoning, not lookup alone.
\rev{The exception is BOAD on 35B-A3B, where dense retrieval attains both a higher pass rate and lower energy.}

\begin{table}[t!]
\centering
\setlength{\tabcolsep}{5pt}
\renewcommand{\arraystretch}{1.15}
\resizebox{\linewidth}{!}{
\begin{tabular}{lrrr}
\toprule
Method & Pass (\%) & Out tok (K) & Energy (kJ) \\
\midrule
\rowcolor{gray!12}
\multicolumn{4}{l}{\textbf{Qwen3.6-35B-A3B}} \\
HyperAgent & 71.0 & 44.9 & 23.77 \\
\hspace{1ex}$+$ Sparse retrieval & \textbf{72.0} & 44.4 & 23.65 \\
\hspace{1ex}$+$ Dense retrieval & 69.0 & 47.0 & 25.05 \\
\hspace{1ex}$+$ Librarian & \textbf{72.0} & \textbf{39.5} & \textbf{19.75} \\
\midrule
BOAD & \textbf{77.0} & 28.6 & 17.47 \\
\hspace{1ex}$+$ Sparse retrieval & 76.0 & 22.3 & 12.93 \\
\hspace{1ex}$+$ Dense retrieval & \textbf{77.0} & \textbf{21.4} & \textbf{12.42} \\
\hspace{1ex}$+$ Librarian & 74.0 & 22.8 & 13.61 \\
\midrule
\rowcolor{gray!12}
\multicolumn{4}{l}{\textbf{Qwen3.6-27B}} \\
HyperAgent & \textbf{73.0} & 33.1 & 114.86 \\
\hspace{1ex}$+$ Sparse retrieval & 70.0 & 28.0 & 97.95 \\
\hspace{1ex}$+$ Dense retrieval & 69.0 & 30.4 & 107.11 \\
\hspace{1ex}$+$ Librarian & 70.0 & \textbf{23.6} & \textbf{82.42} \\
\midrule
BOAD & \textbf{76.0} & 23.0 & 84.25 \\
\hspace{1ex}$+$ Sparse retrieval & 68.0 & 20.0 & 72.86 \\
\hspace{1ex}$+$ Dense retrieval & 74.0 & 20.1 & 73.36 \\
\hspace{1ex}$+$ Librarian & 74.0 & \textbf{19.4} & \textbf{71.08} \\
\bottomrule
\end{tabular}
}

\caption{Comparison between the Librarian and the retrieval-based context management baselines. We compare the Librarian against two baselines that swap the Librarian sub-agent's persistent-history context for one constructed by retrieval. We report the per-episode means on 100 randomly sampled tasks per backbone. \rev{The Librarian attains the lowest energy at a pass rate at or above both retrieval baselines in 3 of the 4 settings.}}
\label{tab:retrieval-comparison}
\end{table}

\section{Conclusion}

In this work, we identify and eliminate a major source of energy waste in multi-agent SWE systems. We first analyze per-turn energy across token categories and find that output tokens dominate. Tracing the excess output, we then identify re-exploration of the same files across invocations as the primary source. We finally address this with the \Librarian, a sub-agent that retains its exploration history across invocations. Integrating the \Librarian into existing MAS reduces per-episode GPU energy while preserving task performance. Ablations attribute the reduction to all four of its mechanisms, with pointer-only answers and the lookup-only scope contributing most on both systems.

\newpage
\section*{Limitations}

In this work, we propose a sub-agent which reduces the energy of the existing multi-agent SWE systems. 
We acknowledge two limitations of our current study.
First, the solution is limited to code navigation. 
The executor sub-agent in HyperAgent, which reproduces the issue and runs tests, also produces duplicate outputs across invocations. Extending the \Librarian's core design to such execution-oriented roles is a promising next step toward system-wide energy reduction.
Second, the scale of the open LLMs we evaluate is limited. Our experiments cover backbones in the active-parameter range of roughly 3B to 27B, and it remains open whether the observed reductions in output tokens and energy carry over to substantially larger open models over 500B inference parameters.

\section*{Acknowledgments}

This work was supported by the National Research Foundation of Korea (NRF) grant funded by the Korea government (MSIT) (RS-2024-00337955); by the Institute of Information \& Communications Technology Planning \& Evaluation (IITP) grants funded by the Korea government (MSIT) (RS-2024-00457882, Artificial Intelligence Research Hub Project; RS-2019-II191906, Artificial Intelligence Graduate School Program (POSTECH)); and by IITP under the Leading Generative AI Human Resources Development (IITP-2026-RS-2026-25546560) grant funded by the Korea government (MSIT).



\bibliography{custom}

\appendix

\section{Per-Token Energy Regression: Implementation Details}
\label{sec:energy-diagnostics}

This appendix details the measurement and fitting behind the energy attribution of \S\ref{sec:energy-characterization}. We first formalize the per-turn token counts and GPU-energy measurement introduced in \S\ref{sec:preliminaries} (Appendix~\ref{sec:per-turn-statistics}), then specify the linear regression and reliability diagnostics that produce Table~\ref{tab:per-token-energy} and Table~\ref{tab:energy-diag} (Appendix~\ref{sec:linear-regression-energy}).

\subsection{Formalism of per-turn statistics}
\label{sec:per-turn-statistics}

For each turn we record two quantities that the regression of Appendix~\ref{sec:linear-regression-energy} later relates: the three token counts the call processes and the GPU energy it draws. We formalize each in turn.

\paragraph{Token counts.}
Each turn's LLM call processes an input context $c_t$ and generates a sequence of output tokens, and its energy cost depends on how these tokens split across three inference operations: auto-regressive decoding, key-value cache lookup, and prefill~\citep{niu2026tokenpowerbench, fernandez-etal-2025-energy}. We therefore record three per-turn token counts. The output count $x^o_t$ is the number of tokens produced by auto-regressive decoding. The input $c_t$ splits further, since modern inference systems reuse the key-value hidden states of recurring prefixes: a cached prefix $x^c_t = |\operatorname{prefix}_{\mathrm{cache}}(c_t)|$ is served by key-value cache lookup, while the remaining $x^u_t = |c_t| - x^c_t$ tokens are processed by prefill.

\paragraph{GPU energy.}
GPU energy is the time integral of device power, from which we subtract an idle baseline to isolate the dynamic draw of active computation.
Let $\tau^{\mathrm{start}}_t$ and $\tau^{\mathrm{end}}_t$ be the wall-clock times bracketing the LLM call of turn $t$, and $\Delta\tau_t = \tau^{\mathrm{end}}_t - \tau^{\mathrm{start}}_t$ its duration. Given the instantaneous GPU power $P(\tau)$ and the idle power $P_{\mathrm{idle},t}$ during the turn, the idle-subtracted dynamic energy of turn $t$ is
\[
\begin{aligned}
E^{\mathrm{net}}_t &= \int_{\tau^{\mathrm{start}}_t}^{\tau^{\mathrm{end}}_t} \left(P(\tau) - P_{\mathrm{idle},t}\right) d\tau \\
&= \int_{\tau^{\mathrm{start}}_t}^{\tau^{\mathrm{end}}_t} P(\tau)\, d\tau - P_{\mathrm{idle},t}\,\Delta\tau_t.
\end{aligned}
\]
The integral term is the cumulative GPU energy drawn over the turn,\footnote{NVIDIA's NVML exposes this quantity as a hardware-level counter.} so $E^{\mathrm{net}}_t$ follows from reading the counter at $\tau^{\mathrm{start}}_t$ and $\tau^{\mathrm{end}}_t$ and subtracting the idle term $P_{\mathrm{idle},t}\,\Delta\tau_t$.

\subsection{Linear regression over token categories}
\label{sec:linear-regression-energy}

We fit the per-turn energy model of \S\ref{sec:token-level-energy-attribution} by ordinary least squares and verify that its coefficients are reliable, describing the fitted data, the regression specification, and the reliability diagnostics in turn.

\subsubsection{Setup}
\label{sec:energy-data}
We run the three systems studied in \S\ref{sec:token-level-energy-attribution}, namely the single-agent Vanilla and the MAS baselines BOAD and HyperAgent, on SWE-Bench Verified, with Qwen3.6-35B-A3B on both A100 and B200 and Qwen3.6-27B on B200. The sample sizes used in Table~\ref{tab:per-token-energy} are 23{,}353 turns / 185 episodes (A3B/A100), 196{,}310 turns / 1{,}500 episodes (A3B/B200), and 179{,}011 turns / 1{,}500 episodes (27B/B200).

\subsubsection{Regression model}
\label{sec:energy-model-spec}

The dependent variable for turn $t$ is the idle-subtracted dynamic energy
\begin{equation}
\label{eq:e-net}
E^{\mathrm{net}}_t \;=\; E^{\mathrm{raw}}_t \;-\; P_{\mathrm{idle},t}\,\Delta\tau_t,
\end{equation}
where $E^{\mathrm{raw}}_t$ is the GPU energy integrated over the call, $P_{\mathrm{idle},t}$ the NVML idle power reading, and $\Delta\tau_t$ the call duration, all in millijoules. With design matrix $X = [\,\mathbf{1}\;\;X^u\;\;X^c\;\;X^o\,] \in \mathbb{R}^{n \times 4}$, the per-turn model reproduces the regression of \S\ref{sec:token-level-energy-attribution},
\begin{equation*}
E^{\mathrm{net}}_t \;=\; \alpha + \beta_u\,x^u_t + \beta_c\,x^c_t + \beta_o\,x^o_t + \epsilon_t,
\end{equation*}
where $\alpha$ captures fixed per-turn overhead and is omitted from Table~\ref{tab:per-token-energy}. Collecting the per-turn responses $E^{\mathrm{net}}_t$ into the vector $y$, we fit by ordinary least squares,
\begin{equation}
\label{eq:ols-normal}
\hat\beta \;=\; \arg\min_{b \in \mathbb{R}^4} \|y - X b\|_2^2 \;=\; (X^\top X)^{-1} X^\top y,
\end{equation}
using \texttt{np.linalg.lstsq} without centering, scaling, regularization, or robust losses.

\subsubsection{Reliability diagnostics}
\label{sec:energy-diag-section}

Table~\ref{tab:energy-diag} reports three diagnostics per fit, namely the overall $R^2$, the variance-inflation factor per coefficient, and the partial $R^2$ per coefficient.

\paragraph{Overall $R^2$.}
$R^2$ on the fit sample is the conventional in-sample fraction of variance explained,
\begin{equation*}
R^2 \;=\; 1 \;-\; \frac{\sum_t (y_t - \hat y_t)^2}{\sum_t (y_t - \bar y)^2}, \qquad \hat y = X\hat\beta.
\end{equation*}

\paragraph{Variance-inflation factor.}
VIF per regressor $j \in \{u, c, o\}$ measures how well column $X_j$ is linearly explained by the remaining regressors. Let $R^2_j$ be the $R^2$ of regressing $X_j$ on $X_{-j}$, fitted with the same \texttt{lstsq} call. The variance-inflation factor is then
\begin{equation*}
\mathrm{VIF}_j \;=\; \frac{1}{1 - R^2_j}.
\end{equation*}
Values close to 1 indicate near-orthogonality with the other regressors.

\paragraph{Partial $R^2$.}
The partial $R^2$ for coefficient $j$ is the share of variance in $y$ uniquely explained by $X_j$ after $X_{-j}$ is partialled out, and we compute it via Frisch--Waugh--Lovell residualisation. Let $M_{-j} = I - X_{-j}(X_{-j}^\top X_{-j})^{-1} X_{-j}^\top$ be the residual-projector onto the orthogonal complement of $X_{-j}$, and write $e_y = M_{-j} y$ and $e_{X_j} = M_{-j} X_j$ for the corresponding residuals. The partial $R^2$ is then the $R^2$ of the univariate regression of $e_y$ on $e_{X_j}$,
\begin{equation*}
\mathrm{partial}\,R^2_j \;=\; \frac{\bigl(e_y^\top e_{X_j}\bigr)^2}
                                    {\bigl(e_{X_j}^\top e_{X_j}\bigr)\,
                                     \bigl(e_y^\top e_y\bigr)}.
\end{equation*}

\begin{table}[t!]
\centering
\resizebox{\columnwidth}{!}{
\begin{tabular}{llrrr}
\toprule
Setting & Coefficient & Pearson $r$ & VIF & partial $R^2$ \\
\midrule
\multirow{3}{*}{A3B / A100 (R$^2$=0.979)} & $\beta_u$ & $0.157$ & $1.01$ & $0.4128$ \\
 & $\beta_c$ & $0.228$ & $1.02$ & $0.3064$ \\
 & $\beta_o$ & $0.978$ & $1.02$ & $0.9769$ \\
\midrule
\multirow{3}{*}{A3B / B200 (R$^2$=0.987)} & $\beta_u$ & $0.050$ & $1.01$ & $0.0705$ \\
 & $\beta_c$ & $0.111$ & $1.02$ & $0.0584$ \\
 & $\beta_o$ & $0.993$ & $1.01$ & $0.9870$ \\
\midrule
\multirow{3}{*}{27B / B200 (R$^2$=0.992)} & $\beta_u$ & $0.097$ & $1.04$ & $0.1806$ \\
 & $\beta_c$ & $0.133$ & $1.05$ & $0.1877$ \\
 & $\beta_o$ & $0.995$ & $1.02$ & $0.9919$ \\
\bottomrule
\end{tabular}
}

\caption{Coefficient reliability for Table~\ref{tab:per-token-energy}. Pearson $r$ is the univariate correlation between each regressor and turn energy. VIF denotes the variance-inflation factor, where values close to 1 indicate no collinearity with the other regressors. Partial $R^2$ is the Frisch--Waugh semi-partial $R^2$, namely the share of variance in turn energy uniquely explained by this coefficient after the others are partialled out. All three setups satisfy $\mathrm{VIF} \le 1.05$ and $R^2 \ge 0.97$.}
\label{tab:energy-diag}
\end{table}

\section{Librarian Implementation Details}
\label{app:librarian}

This appendix specifies how a \Librarian invocation manages and returns its context, then reproduces every prompt it uses. Two mechanisms keep the persistent context both current and compact across invocations---history pruning (Appendix~\ref{app:librarian:pruning}) and the freshness report (Appendix~\ref{app:librarian:freshness})---and the submission step (Appendix~\ref{app:librarian:submission}) turns the \Librarian's pointer-only answer into the verbatim excerpts the orchestrator receives. We then reproduce every prompt used by the \Librarian and by the per-MAS integration recipe, \rev{and Appendix~\ref{app:librarian:noscope} specifies how the lookup-only scope ablation of \S\ref{sec:exp-ablation} is constructed, that mechanism having no canonical null form.}

\subsection{History pruning}
\label{app:librarian:pruning}

The \Librarian persists its full trajectory across invocations within an episode, but keeps an invocation only if that invocation surfaced genuinely new code. After an invocation returns, we estimate the previously-unseen file content it produced and discard the invocation when that estimate is small.

\paragraph{Detecting file reads.} A turn is counted as a file read if it either (i)~calls the OpenHands \texttt{str\_replace\_editor} tool in \texttt{view} mode, or (ii)~runs one of \texttt{cat}, \texttt{sed\,-n}, \texttt{head}, \texttt{tail}, or \texttt{awk} on a repository file. From the tool call we recover the target file and the line range it reads; each such read returns a line-numbered observation of that range.

\paragraph{Counting new content.} Novelty is judged against the \Librarian's \emph{entire retained session}, not the current invocation alone: the \Librarian keeps a single cumulative record of which lines of each file it has already viewed across all kept invocations, and a read counts only the observation characters on lines absent from that record. A read confined to already-seen lines therefore contributes zero, a partially overlapping read is scaled by the fraction of newly-seen lines, and a whole-file view counts only on first sight; the newly read lines are then added to the record. Two rules keep this record consistent with the retained history: a pruned invocation (below) has its additions rolled back, and the freshness report clears a file's entry whenever that file is written, so a re-read of its changed content counts as new again. Summing over the invocation gives its new-content size.

\paragraph{Pruning rule.} If the new-content size is below $500$ characters, the invocation is treated as contributing nothing reusable to future queries and its turns are discarded---the next invocation's context is reconstructed as if it never occurred. Otherwise its turns are appended to the persistent history. Trivial or purely re-derived lookups thus leave no residue, while substantive lookups accumulate.

\subsection{Freshness report}
\label{app:librarian:freshness}

Because the editor and executor may modify files between two \Librarian invocations, an excerpt cached in the persistent history can go stale. Every invocation with non-empty history therefore prepends a \emph{freshness report} that tells the \Librarian which cached regions it may still trust. The report is built in two steps.

\paragraph{Modified ranges.} We run \texttt{git diff -U0 HEAD} in the repository and read each file's modified line ranges directly from the hunk headers. Every line location reported to the \Librarian is derived from this diff.

\paragraph{Selecting which files to surface.} The \Librarian stores a hash of each file's \texttt{git diff} section. At every invocation it re-runs \texttt{git diff} and, for any file whose hash is new or differs from the stored one, adds the file to the report, updates the stored hash, and resets that file's viewed-line cache; files whose hash is unchanged are omitted.

\paragraph{Contents.} For each surfaced file the report lists its modified line ranges as pointers (e.g.\ \texttt{path: lines [10--15, 40]}), not the diff text. The accompanying instruction states that content outside those ranges is unchanged and should be quoted from history, and that a listed range need be re-viewed only if a prior view intersects it; reverted files are flagged so their cached excerpts are discarded. When no file has changed since the previous invocation, the report states this and directs the \Librarian to answer from history. The example below is drawn from an episode on the \texttt{astropy} repository, where the intervening edits touched three line ranges of \texttt{sliced\_wcs.py}.

\begin{promptbox}{Example file freshness report}
\begin{Verbatim}
[FILE FRESHNESS REPORT -- NEW edits since the previous wake. For each listed file, content OUTSIDE these ranges is unchanged from when you last viewed it -- quote from history. Re-view ONLY if your prior view intersects the listed ranges. Files NOT listed here are either unchanged entirely or already known-modified from a prior freshness report.]
  - astropy/wcs/wcsapi/wrappers/sliced_wcs.py: lines [188-196, 258, 264-265]
\end{Verbatim}
\end{promptbox}

\subsection{Pointer-only submission}
\label{app:librarian:submission}

A \Librarian invocation ends by calling \texttt{submit}, the tool that realizes the pointer-only answers of \S\ref{sec:method:invocation}. The call carries two fields: \texttt{result}, a short natural-language description of what was found, and \texttt{view\_commands}, a list of \texttt{[path, start, end]} triples naming the located line ranges. The code itself is referenced through these pointers and never inlined into \texttt{result}, so the tokens the \Librarian generates stay bounded by the length of its description, however much code it surfaces.

\paragraph{Submit call.} The example below is from an episode on the \texttt{django} repository, where the orchestrator asked where a management command initializes its output streams. The \Librarian locates the relevant range and submits one view command alongside a one-sentence summary.

\begin{promptbox}{Example Librarian submission}
\begin{Verbatim}
submit({"result": "Lines 230-260 of `/testbed/django/core/management/base.py` shown below. Stdout/stderr initialization is in `__init__` at lines 239-248 — `self.stdout` and `self.stderr` are `OutputWrapper` instances wrapping `sys.stdout`/`sys.stderr`. `stderr.style_func` is set to `self.style.ERROR` for error coloring.", "view_commands": "[[\"/testbed/django/core/management/base.py\", 230, 260]]"})
\end{Verbatim}
\end{promptbox}

\paragraph{Transformation into the orchestrator message.} The tool executes each view command on the \Librarian's behalf---equivalent to \texttt{cat\,-n} over the named range---and builds the orchestrator-facing message by prepending the resulting line-numbered excerpts to the \texttt{result} text, tagging the whole with the delegation tool name (here \texttt{delegate\_to\_librarian}). The orchestrator thus receives the verbatim code that the \Librarian never emitted, followed by the \Librarian's summary, while the submission the \Librarian generated remained the short pointer list above.

\begin{promptbox}{Example message delivered to the orchestrator}
\begin{Verbatim}
[delegate_to_librarian] Here's the result of running `cat -n` on /testbed/django/core/management/base.py lines 230-260:
   230        output_transaction = False  # Whether to wrap the output in a "BEGIN; COMMIT;"
   231        requires_migrations_checks = False
   232        requires_system_checks = '__all__'
   233        # Arguments, common to all commands, which aren't defined by the argument
   234        # parser.
   235        base_stealth_options = ('stderr', 'stdout')
   236        # Command-specific options not defined by the argument parser.
   237        stealth_options = ()
   238    
   239        def __init__(self, stdout=None, stderr=None, no_color=False, force_color=False):
   240            self.stdout = OutputWrapper(stdout or sys.stdout)
   241            self.stderr = OutputWrapper(stderr or sys.stderr)
   242            if no_color and force_color:
   243                raise CommandError("'no_color' and 'force_color' can't be used together.")
   244            if no_color:
   245                self.style = no_style()
   246            else:
   247                self.style = color_style(force_color)
   248                self.stderr.style_func = self.style.ERROR
   249            if self.requires_system_checks in [False, True]:
   250                warnings.warn(
   251                    "Using a boolean value for requires_system_checks is "
   252                    "deprecated. Use '__all__' instead of True, and [] (an empty "
   253                    "list) instead of False.",
   254                    RemovedInDjango41Warning,
   255                )
   256                self.requires_system_checks = ALL_CHECKS if self.requires_system_checks else []
   257            if (
   258                not isinstance(self.requires_system_checks, (list, tuple)) and
   259                self.requires_system_checks != ALL_CHECKS
   260            ):

Lines 230-260 of `/testbed/django/core/management/base.py` shown below. Stdout/stderr initialization is in `__init__` at lines 239-248 — `self.stdout` and `self.stderr` are `OutputWrapper` instances wrapping `sys.stdout`/`sys.stderr`. `stderr.style_func` is set to `self.style.ERROR` for error coloring.
\end{Verbatim}
\end{promptbox}

\subsection{Librarian system prompt}
\label{app:librarian:prompt}

The full system prompt for the \Librarian sub-agent. It restricts the role to locating code and returning verbatim excerpts, enumerates the in-scope and out-of-scope behaviors, defines the submission interface, and specifies the pre-submit self-check that scrubs analytical phrasing from the prose answer.

\begin{promptbox}{Prompt for the Librarian sub-agent}
\begin{Verbatim}
You are the repository Librarian. Your role is strictly limited to **locating files / code regions and returning verbatim excerpts**. You are NOT an analyst, debugger, or issue interpreter -- those are the main agent's job.

Scope (what you DO):
  - Locate the file path + line range where a named symbol (class, function, method, constant, fixture, test) lives.
  - Return verbatim code excerpts at requested locations.
  - Surface adjacent context (callers, sibling tests, related fixtures) when the asker is clearly going to need it.
  - List directory layouts when asked.

Out of scope (what you DO NOT do -- even if the main agent asks):
  - Do NOT analyse why code does what it does, infer intent, or explain bugs / fix logic.
  - Do NOT interpret git history / commit messages / past PRs. Specifically, do NOT make claims like "this code is a fix for issue #N" or "this was added to handle X" -- those are historical inferences you cannot ground in the current code.
  - Do NOT propose fixes or describe what the main agent should edit. If asked, reply with the code at the location and stop.
  - Do NOT translate an abstract issue description into "the bug is at line N because Y". Just return the code at the locations the main agent named (or the closest match) and let them reason.
  When the asker requests analysis (e.g. "what's the issue here?", "how does this fix the bug?"), you reply with the relevant code excerpts and a one-line note like "Code at <path>:<lines> is shown above; analysis is the main agent's responsibility."

Tools available to you:
  - ``bash`` (view-only -- ``find``, ``grep``, ``rg``, ``ls``, ``tree``, ``wc``, ``cat`` of short files, etc. No edits, no test runs, no destructive commands.)
  - ``str_replace_editor`` (view-only -- ``view`` command only, for inspecting file ranges. No create/edit.)
  - ``submit(result, view_commands)`` -- call this exactly once per consultation to return your findings to the main agent and end the call. ``result`` is short prose; ``view_commands`` is an optional JSON list of ``[path, start, end]`` triples. See "Reply SHAPE" below for the contract.

Persistence semantics (the "memory" half of your role): You are the SAME Librarian instance across every ``repo_librarian`` call in this episode. Your full conversation history is visible when you wake up -- prior tool outputs, prior file excerpts, prior ``submit`` summaries, all of it. BEFORE issuing a new bash / view command, scan your own history: if the answer is already there, do NOT re-read; quote it directly. When you do run a new command, pick one that strictly extends what you already have rather than re-fetching it.

Handling ambiguous queries (this is real -- the main agent sometimes sends keyword-soup queries with no clear question):
  - First try to disambiguate from your own accumulated history and one or two well-targeted probes. If the keywords still resolve to multiple plausible interpretations, do NOT keep grinding more grep variants hoping one lands. Stop and submit a clarifying reply via ``submit`` instead -- list the candidate interpretations you considered (with the file/symbol each one points at) and ask which one the agent meant.
  - This is preferable to a long grep loop because: (a) the next consultation will arrive with a sharper question, (b) your history retains the candidates you already enumerated, so the follow-up is cheap.
  - Clarification replies are SHORT -- the candidate list, file pointers if you have them, and the question. Do NOT include code excerpts on a clarification.

Reply discipline (when the query is well-formed):
  - Ground every claim in a file path + line range from your tool output (or your prior history).
  - When the main agent's query points at a behaviour or symptom, translate it into "what file / symbol owns this surface area" and reply with that location. Do NOT volunteer a diagnosis of the symptom -- that's the main agent's job.

Reply SHAPE -- describe locations in ``result``, attach code via ``view_commands``:

The ``submit`` tool takes TWO parameters:
  - ``result``: short prose answer (<~ 200 tokens). File paths + line ranges + brief framing of what each region is. **Do NOT paste verbatim code into ``result``.** Bytes are bigger than your TEXT output should be.
  - ``view_commands``: optional JSON list of ``[path, start, end]`` triples. The system runs each view server-side and prepends the output to what the main agent sees. The view does NOT enter YOUR conversation history (it's executed outside the SDK turn), so your token budget for future ``repo_librarian`` calls in this episode stays lean.

Three response shapes:

  - **Pointer (default)**: short ``result`` describing the location, single-element ``view_commands`` with the region you're pointing at. Example::

        result = (
          "Found `rotation_matrix` in "
          "/testbed/astropy/coordinates/matrix_utilities.py "
          "lines 30-78 -- body included below."
        )
        view_commands = '[["/testbed/astropy/coordinates/matrix_utilities.py", 30, 78]]'

    Pick the line range carefully so it lands on natural boundaries (class header -> next sibling, function header -> matching dedent).

  - **Multi-fragment**: short ``result`` summarising what you found, ``view_commands`` with multiple regions. Use ONLY when:
      * each fragment is <= ~15 lines, AND
      * the fragments span >= 2 different files OR >= 2 non-adjacent regions of one file, AND
      * the main agent obviously needs ALL of them to plan their next move.
    Example::

        result = (
          "Three call sites of `where_not_allclose`: one in diff.py "
          "table comparison, two in nddata arithmetic. All quoted below."
        )
        view_commands = (
          '[["/testbed/astropy/io/fits/diff.py", 1444, 1471],'
          ' ["/testbed/astropy/nddata/mixins/ndarithmetic.py", 220, 230],'
          ' ["/testbed/astropy/nddata/mixins/ndarithmetic.py", 410, 425]]'
        )

  - **Clarification**: ``result`` lists candidate interpretations and file pointers; ``view_commands`` omitted (or ``None``). Use when the query is genuinely ambiguous.

If the answer is fully reconstructable from your prior history without new tool calls, you may still call ``submit`` with ``view_commands`` pointing at locations you already viewed earlier in this episode -- the system re-fetches them, you don't have to re-issue ``view`` first.

Pre-submit self-check (apply BEFORE every ``submit`` call):
  1. Is ``result`` PROSE only (no ``def ...`` / ``class ...`` / line-numbered code blocks)? If I caught myself pasting code into ``result``, move it to ``view_commands`` instead.
  2. Are the line ranges in ``view_commands`` PRECISE (verified against my actual ``view`` / ``cat -n`` output)? Vague ranges like ``[200, 300]`` make the main agent see the wrong region -- re-ground from output.
  3. For multi-fragment ``view_commands``: are the regions REALLY scattered (different files OR non-adjacent regions), or am I really showing one long block split into "fragments"? If the latter, collapse to a single ``[path, start, end]`` covering the union range.
  4. Did my ``result`` slip into analysis / interpretation territory?  Phrases to scrub before submitting:
       - "this is a fix for issue #N" / "this was added to handle X" (historical inference -- not your job)
       - "the bug is here because ..." / "the issue arises when ..." (diagnosis -- not your job)
       - "the fix should ..." / "you should change this to ..." (proposal -- not your job)
     If you find any, delete them. Keep file:line framing only.
\end{Verbatim}
\end{promptbox}

\subsection{Tool-selection policy}
\label{app:librarian:tool-policy}

The block spliced into each MAS's orchestrator prompt. It instructs the orchestrator to route unknown-location queries to the \Librarian and to reuse a \texttt{file:line} pointer it already holds instead of consulting the \Librarian a second time. \rev{The block takes two forms, selected by whether the orchestrator can read files itself. Both carry the same decision rule, the same persistence note, the same continuous-consultation policy, and the same query-phrasing rules; they differ in how the orchestrator acts on a pointer once it has one.} The placeholder \texttt{\{librarian\_tool\}} is replaced at MAS-instantiation time with the specific delegation tool name.

\paragraph{\rev{Orchestrators with direct file access.}}
\rev{Used by the vanilla MAS, where the tool is named \texttt{repo\_librarian}, and by BOAD, where it is named \texttt{librarian}. Both orchestrators own \texttt{bash} and the file view tool, so the policy tells them to view a known \texttt{file:line} themselves and forbids using \texttt{bash} for symbol or pattern search.}

\begin{promptbox}{Prompt for the tool-selection policy of orchestrators with direct file access}
\begin{Verbatim}
Tool selection policy (read this BEFORE you plan):
  - **Decision rule**: if you ALREADY know the precise file path AND a useful line range to inspect, ``str_replace_editor view`` it directly. If you DON'T know where to look -- you have a symbol/behaviour/symptom but no file:line -- ask ``{librarian_tool}``.
  - ``{librarian_tool}`` is your **search** tool -- it returns file paths and line ranges (and short excerpts when several fragments are scattered). Treat it as a sharper, memory-equipped grep, NOT as an analyst. Use it to ask "where is X?" / "which file owns Y?" / "give me the file:line for Z" -- never "what's the bug?" or "is this code correct?". Reasoning, diagnosis, and fix-design stay with you.
  - The librarian is a single long-lived instance: its conversation history accumulates across calls, so follow-ups are cheap.
  - ``bash`` and ``str_replace_editor`` are for both ACTING on the code (editing source, running tests, installing dependencies) AND READING code at known locations (``str_replace_editor view`` with a file path + line range you already have). What ``bash`` is NOT is a search tool.
  - Continuous consultation policy: each NEW search question (a symbol you don't yet have a location for, a "where does X live?" question) is a NEW librarian call. Once the librarian has given you a path + line range, you own that location -- viewing nearby lines, the enclosing class, or an adjacent helper at a known file:line is your job, NOT another librarian call. The single biggest failure mode in BOTH directions: (a) calling the librarian once and then doing keyword grep yourself for every subsequent symbol -- do NOT do that; (b) re-asking the librarian for a region you already have a precise file:line for -- do NOT do that either, just ``view`` it.
  - Forbidden in ``bash``: ``find`` / ``grep`` / ``rg`` / ``ls`` / ``tree`` / ``sed`` / ``awk`` / ``less`` / ``more`` for symbol or pattern search across source files. (Reading a specific file you already located via ``cat`` / ``head`` / ``tail`` is fine but ``str_replace_editor view`` with ``view_range`` is preferred -- it gives you line numbers.) Searching for unknown symbols is the librarian's job.
  - Query phrasing: write each ``{librarian_tool}`` query as a **lookup question**, not a bag of keywords and not an open-ended analysis request. Good shapes: "Where is <symbol> defined? I need the file path + line range.", "Which file owns <behaviour> in <surface area>?", "Locate <test name>'s file path.". Bad shapes: a keyword string with no question, or an analysis prompt like "Why does <symptom> happen?" / "What is the bug here?" -- the librarian will only return locations, not analysis.
  - Healthy turn pattern: (don't know where) librarian call -> librarian returns file:line pointer -> you ``str_replace_editor view`` to read -> reason -> edit -> run tests -> (need a NEW location? librarian again).
\end{Verbatim}
\end{promptbox}

\paragraph{\rev{Orchestrators that act by delegation.}}
\rev{Used by HyperAgent, where the tool is named \texttt{delegate\_to\_librarian}. Its planner holds no file tool at all, so the two bullets above are replaced by their delegation analogue: the planner must spell the \Librarian's pointer out in the request it sends to the editor or executor, so that neither of them searches for the location again.}

\begin{promptbox}{Prompt for the tool-selection policy of delegation-only orchestrators}
\begin{Verbatim}
Tool selection policy (read this BEFORE you delegate):
  - **Decision rule**: when you don't know WHERE in the repo something lives (a symbol, behaviour, or symptom but no file:line), ask ``{librarian_tool}``. When you HAVE a precise file:line and want code changed or tests run, ask ``delegate_to_editor`` / ``delegate_to_executor`` with the file:line spelled out explicitly in your ``request`` -- do NOT make them search for it again.
  - ``{librarian_tool}`` is your **search** intern -- it returns file paths and line ranges (and short excerpts when several fragments are scattered). Treat it as a sharper, memory-equipped grep, NOT as an analyst. Use it to ask "where is X?" / "which file owns Y?" / "give me the file:line for Z" -- never "what's the bug?" or "is this code correct?". Reasoning, diagnosis, and fix-design stay with you.
  - The librarian is a single long-lived instance: its conversation history accumulates across calls, so follow-ups (callers of a symbol, subclasses that override it, adjacent helpers, sibling tests) are cheap.
  - You have NO direct view tool. Once the librarian has given you a precise file:line, *you own that location* -- when you next delegate to the editor or executor, name the explicit path + line range in your ``request`` so they can read or edit directly without re-searching.
  - Continuous consultation policy: each NEW search question (a symbol you don't yet have a location for, a "where does X live?" question) is a NEW librarian call. Once the librarian has given you a path + line range, do NOT re-ask the librarian for that region. The two biggest failure modes here: (a) you call the librarian once for one location, then delegate "find X and fix it" to the editor for the next location -- DO NOT let the editor or executor re-search; route every new "where does X live?" through the librarian and pass the answer down; (b) you re-ask the librarian for a region the librarian has already pinpointed earlier in this episode -- just reuse what's in its accumulated history (the librarian will quote it back cheaply).
  - Query phrasing for the librarian: write each ``{librarian_tool}`` ``request`` as a **lookup question**, not a bag of keywords, not "Show me the contents of <file>", and not an open-ended analysis request. Good shapes: "Where is <symbol> defined? I need the file path + line range.", "Which file owns <behaviour> in <surface area>?", "Locate <test name>'s file path.". Bad shapes: a keyword string with no question, "Show me the contents of <file>" (the librarian is not a paste service -- name the symbol/region you actually need), or an analysis prompt like "Why does <symptom> happen?" / "What is the bug here?" -- the librarian only returns locations, not analysis.
  - Healthy turn pattern: (don't know where) ``{librarian_tool}`` -> librarian returns file:line pointer -> ``delegate_to_editor`` with that file:line for the change OR ``delegate_to_executor`` with that file:line in the test command -> (need a NEW location? librarian again).
\end{Verbatim}
\end{promptbox}

\subsection{Plan rewriting guideline}
\label{app:librarian:rewrite-guide}

The instructions we follow to derive each $+$\Librarian plan from its baseline. The guideline encodes two rules, namely renaming references to the prior code-navigator sub-agent and rewriting any code or file location step to a \Librarian delegation call. The same guideline applies across MAS, with the actual scope of changes varying by how much navigation each MAS previously routed through bash versus delegation.

\begin{promptbox}{Prompt for the plan-rewriting guideline}
\begin{Verbatim}
# Plan / Planner Prompt Rewriter (-> +librarian variant)

You are rewriting a baseline MAS orchestrator / planner prompt into its **+librarian** variant. The transform replaces a navigator-style sub-agent with the persistent Repo Librarian, and rewrites the plan steps that involve code / file location so the lookup goes through the librarian.

You edit ONLY the orchestrator/planner-facing prompt. Other sub-agents' prompts are out of scope.

---

## INPUT FORMAT

You receive four sections, separated by tagged delimiters:

    <previous_subagent_name>
    ... name of the navigator-style sub-agent being replaced (e.g. "code_navigator", "Codebase Navigator", "delegate_to_navigator") ...
    </previous_subagent_name>

    <previous_subagent_prompt>
    ... the full system prompt of the previous sub-agent, used to identify which of its responsibilities go beyond file/code lookup and therefore need to be redirected rather than absorbed by the librarian ...
    </previous_subagent_prompt>

    <baseline_prompt>
    ... the baseline plan / planner prompt to rewrite ...
    </baseline_prompt>

    <librarian_spec>
    ... the librarian's contract: its role, persistence semantics, query-shape rules, and what it does / does not do ...
    </librarian_spec>

---

## RULES

### 1. Replace the named sub-agent with the Librarian

Every reference to the sub-agent name(s) in `<previous_subagent_name>` -- docstring, intern-list entry, delegation-tool name, inline mention in a numbered step -- becomes the librarian. Use a surface name that matches the baseline's own casing and tool-naming convention (`librarian`, `Repo Librarian`, `delegate_to_librarian`, etc.). Do not paraphrase any other text on the same line.

### 2. Rewrite code/file-location steps to use the Librarian

Any step in the plan that asks the host to **locate, navigate, search, explore, or look up** code (find a symbol, identify affected files, map the codebase, locate call sites, find the test that exercises a behaviour) -- including post-failure debugging steps that re-locate code on a failing test -- gets rewritten so the lookup happens through the librarian.

Preserve the librarian's contract exactly as stated in `<librarian_spec>`: natural-language lookup questions, persistent conversation history across calls within the episode, NEW search question = NEW librarian call, host owns the location once handed back, librarian only LOCATES (host reasons / diagnoses), no fallback to `grep` / `find` for symbol search.

Steps that do **not** involve code/file location (issue analysis, synthesis, fix implementation, cleanup, submit, etc.) stay verbatim.

### 3. Redirect responsibilities that go beyond file/code lookup

Read `<previous_subagent_prompt>` to enumerate the responsibilities the previous sub-agent held beyond pure file/code lookup -- e.g. analyzing code purposes, mapping inter-module dependencies, summarizing architecture, characterizing call relationships. For each such responsibility, do NOT route it to the librarian (the librarian only LOCATES). Instead, fold it into the orchestrator's own analytical steps or into another sub-agent that the baseline plan already invokes for analysis/synthesis. If neither already covers it, add a short orchestrator-side step adjacent to the librarian call that performs that work on the returned excerpts. Leave the librarian call itself unchanged.

---

## BREVITY

Every character is prefilled on every planner / orchestrator turn. Keep the rewrite within ~1.15x the baseline length.

---

## OUTPUT

Return ONLY the rewritten prompt. Preserve the original formatting, section headers, bullet structure, and any verbatim blocks unaffected by the two rules. Do not add commentary before or after.
\end{Verbatim}
\end{promptbox}

\subsection{Per-MAS plan rewrites}
\label{app:librarian:plans}

The orchestrator-side planning sequences for the BOAD and HyperAgent MAS. For each MAS we reproduce the baseline plan alongside the $+$\Librarian variant produced by applying the guideline in Appendix~\ref{app:librarian:rewrite-guide}, with BOAD's variant generated via Claude Opus~4.6.

\paragraph{BOAD baseline.}
The orchestrator's six-step plan with step 2 delegating code mapping to the \texttt{code\_navigator} sub-agent.

\begin{promptbox}{Prompt for BOAD orchestrator plan, baseline}
\begin{Verbatim}
1. Use the issue_analyzer subagent to decompose the problem description into structured requirements, identify affected components, and define explicit success criteria to guide the investigation.

2. Use the code_navigator subagent to map the relevant codebase structure, focusing on files and functions identified in the analysis to understand dependencies and data flow.

3. Synthesize the structured analysis and code mapping to implement the fix, ensuring the solution addresses the root cause while adhering to the established success conditions.

4. Create a minimal test case or validation script to verify the fix against the reproduction criteria defined in the initial analysis.

5. After you have solved the issue, delete any test files or temporary files you created.

6. Use the submit tool to submit the changes to the repository.
\end{Verbatim}
\end{promptbox}

\paragraph{BOAD $+$ Librarian.}
Steps 1, 3, 5, and 6 are byte-identical to the baseline. Step 2 is rewritten to phrase every search as a natural-language lookup question to the \Librarian, and step 4 is rewritten to call the \Librarian for additional location lookups after test failures.

\begin{promptbox}{Prompt for BOAD orchestrator plan with Librarian}
\begin{Verbatim}
1. Use the issue_analyzer subagent to decompose the problem description into structured requirements, identify affected components, and define explicit success criteria to guide the investigation.

2. Use the librarian subagent to map the relevant code areas: phrase each request as a natural-language lookup question ("Where lives <symbol / behaviour>?", "Show me the body of <function>", "Locate the test that exercises <feature>"). The librarian is a single long-lived instance -- its conversation history accumulates across calls in this episode, so follow-up lookups (the caller of a symbol, subclasses that override it, adjacent helpers, sibling tests) are cheaper than starting a fresh search. Each NEW search question (a symbol or behaviour you do not yet have a file:line pointer for) is a NEW librarian call; once the librarian has given you a location, that location is yours -- view nearby lines or the enclosing class yourself rather than re-asking the librarian for it. The librarian only LOCATES code; reasoning about what to change stays with you. Call the librarian aggressively whenever a new location is needed; do not fall back to `grep` / `find` for symbol search. After exploration, propose the files and methods to fix the issue and select the most likely location to apply the change.

3. Synthesize the structured analysis and code mapping to implement the fix, ensuring the solution addresses the root cause while adhering to the established success conditions.

4. Write and execute targeted unit tests or integration checks to validate that the solution resolves the issue without introducing regressions. If any check fails, consult the librarian for ADDITIONAL location lookups (other call sites of a changed symbol, subclasses that override it, fixtures or sibling tests that exercise the failing path) -- phrase each as a location question. The librarian returns a pointer to the relevant code; YOU then view it, diagnose why the test fails, refine your fix, and rerun the affected tests. Do NOT fall back to fresh `grep` / `find` for symbol search -- those go through the librarian. Repeat until all checks pass.

5. After you have solved the issue, delete any test files or temporary files you created.

6. Use the submit tool to submit the changes to the repository.
\end{Verbatim}
\end{promptbox}

\paragraph{HyperAgent baseline.}
The planner's five-step sequence with step 2 delegating to the Codebase Navigator.

\begin{promptbox}{Prompt for HyperAgent planner, baseline}
\begin{Verbatim}
1. Read the PR description and understand the issue.
2. Delegate to the Codebase Navigator to localize where the change must happen.
3. Delegate to the Codebase Editor to apply the minimal source change.
4. Delegate to the Executor to verify the fix (run the relevant tests / a reproduction script).
5. Iterate as needed until the tests pass.
\end{Verbatim}
\end{promptbox}

\paragraph{HyperAgent $+$ Librarian.}
Step 2 is rewritten to delegate to the Repo \Librarian. The remaining four steps are byte-identical to the baseline.

\begin{promptbox}{Prompt for HyperAgent planner with Librarian}
\begin{Verbatim}
1. Read the PR description and understand the issue.
2. Delegate to the Repo Librarian to localize where the change must happen.
3. Delegate to the Codebase Editor to apply the minimal source change.
4. Delegate to the Executor to verify the fix (run the relevant tests / a reproduction script).
5. Iterate as needed until the tests pass.
\end{Verbatim}
\end{promptbox}

\subsection{\rev{Implementation of the lookup-only scope ablation}}
\label{app:librarian:noscope}

\rev{Three of the four mechanisms ablated in \S\ref{sec:exp-ablation} have an obvious null form: the persistent session becomes a fresh sub-agent per invocation, the pointer-only submission becomes one that inlines the code into \texttt{result}, and the freshness report is simply not prepended. The lookup-only scope has no such form, because it restricts what a role is asked to do rather than adding a component that can be switched off. This subsection specifies the variant we use.}

\paragraph{\rev{The counterfactual is the navigator the scope replaced.}}
\rev{Instead of relaxing the restriction by an arbitrary amount, we set the ablated role to the one the lookup-only scope was introduced to displace, namely the code navigator of the host MAS. The \Librarian's system prompt is replaced by HyperAgent's Navigator prompt, which charges the sub-agent with finding all information relevant to the orchestrator's query and imposes no division of labor beyond that. Three elements of the \Librarian prompt therefore disappear: the restriction of the role to locating code, the explicit list of out-of-scope behaviors (analysis, diagnosis, and fix design), and the pre-submit self-check item that scrubs analytical phrasing from the prose answer. BOAD uses the same prompt with every reference to a planner retargeted to the orchestrator, since BOAD has no planner role.}

\paragraph{\rev{The other three mechanisms are held fixed.}}
\rev{The ablated \Librarian keeps its single persistent session, its freshness report, and its pointer-only submission. The last of these requires re-attaching three blocks of the \Librarian prompt to the Navigator base: the reply-shape contract defining \texttt{submit\_result(result, view\_commands)}, the instruction to answer from history whenever the answer is fully reconstructable from it, and the persistence paragraph. The tool set is unchanged as well. The ablated role still holds the view-only \texttt{bash} shell and \texttt{str\_replace\_editor} restricted to \texttt{view}, so removing the scope restores neither editing nor test execution. What the ablation removes is the instructed division of labor, not the sub-agent's capabilities.}

\paragraph{\rev{The orchestrator side is ablated as well.}}
\rev{The scope is a two-sided contract, and \S\ref{sec:method:adapt} installs half of it on the orchestrator. Ablating the sub-agent alone would leave an orchestrator that still refuses to send analytical queries, so we apply the matching edit upstream: the delegation tool is documented as an information gatherer rather than a locator, the tool-selection policy of Appendix~\ref{app:librarian:tool-policy} drops the guidance that rejects open-ended analysis requests, and the rewritten plan steps of Appendix~\ref{app:librarian:plans} are rephrased from lookup questions to information requests. Every other part of the orchestrator prompt, including the SWE context, the user query, and all non-navigation plan steps, stays byte-identical to the full \Librarian.}

\section{Experiment Details}
\label{app:experiment-details}

This appendix records the configuration held fixed across the experiments of \S\ref{sec:experiments}---the vLLM serving, sampling, and scaffold settings (Appendix~\ref{app:hyperparameters}) and the caveman prompting style directive (Appendix~\ref{app:caveman}).

\subsection{Serving and scaffold configuration}
\label{app:hyperparameters}

Table~\ref{tab:hyperparameters} lists the vLLM serving configuration, sampling hyperparameters, and agent scaffold settings used in the experiments of \S\ref{sec:experiments}. The values are held fixed across all backbones and methods so that per-method differences in token consumption reflect orchestration choices rather than serving, decoding, or scaffold variance.

\begin{table}[t!]
\centering
\begin{tabular}{lr}
\toprule
Setting & Value \\
\midrule
\multicolumn{2}{l}{\textit{vLLM serving}} \\
serving context length    & $131{,}072$ \\
GPU memory utilization    & $0.95$ \\
\midrule
\multicolumn{2}{l}{\textit{sampling}} \\
temperature        & $0.6$  \\
top-$p$            & $0.95$ \\
top-$k$            & $20$   \\
min-$p$            & $0.0$  \\
presence penalty   & $0.0$  \\
repetition penalty & $1.0$  \\
\midrule
\multicolumn{2}{l}{\textit{agent scaffold}} \\
shell command timeout (min)         & $30$       \\
tool observation truncation (chars) & $30{,}000$ \\
\bottomrule
\end{tabular}

\caption{vLLM serving configuration, sampling hyperparameters, and agent scaffold settings, held identical across every backbone and method.}
\label{tab:hyperparameters}
\end{table}

\subsection{Caveman prompting directive}
\label{app:caveman}

The fixed style directive that the caveman prompting baseline appends to the system prompts of every sub-agent and the orchestrator. The directive instructs the model to compress its visible answer text while leaving the thinking trace intact.

\begin{promptbox}{Prompt for caveman style directive}
\begin{Verbatim}
OUTPUT STYLE — caveman mode (FULL).

Respond terse like smart caveman. All technical substance stays. Only fluff dies. ACTIVE every response until task done.

Rules:
- Drop articles (a/an/the), filler (just/really/basically/actually/simply), pleasantries (sure/certainly/of course/happy to), hedging.
- Fragments OK. Short synonyms (big not extensive, fix not "implement a solution for").
- Pattern: `[thing] [action] [reason]. [next step].`

NEVER compress / NEVER abbreviate:
- Code blocks, code symbols, function names, class names, file paths, error strings, log lines, shell commands, tool-call arguments.
- Anything inside `backticks`, fenced code, or quoted strings.
- Patch / diff content (must be byte-exact).

Resume normal prose for: security warnings, irreversible-action confirmations, multi-step sequences where fragment order could be misread.

Example
- Normal: "The reason your React component is re-rendering is likely because you're creating a new object reference on each render cycle. You should wrap it in useMemo."
- Caveman: "New object ref each render. Inline object prop = new ref = re-render. Wrap in `useMemo`."
\end{Verbatim}
\end{promptbox}

\section{Full Per-Role Breakdown}
\label{app:role-breakdown}

This appendix reproduces the per-episode breakdown of output tokens and idle-subtracted GPU energy at the granularity of each role inside HyperAgent (Table~\ref{tab:role-comparison-hyperagent}) and BOAD (Table~\ref{tab:role-comparison-boad}). The compact main-text version (Table~\ref{tab:role-comparison-compact}) collapses the non-replaced roles into a single ``other roles'' row by summing within each episode before averaging; the tables below leave every role on its own row so the reader can see the unaggregated contribution of, for example, the HyperAgent executor or the BOAD issue analyzer. 

\begin{table}[t!]
\centering
\setlength{\tabcolsep}{6pt}
\renewcommand{\arraystretch}{1.15}
\resizebox{\linewidth}{!}{
\begin{tabular}{@{}llrrcrr@{}}
\toprule
 & & \multicolumn{2}{c}{HyperAgent} & & \multicolumn{2}{c}{HyperAgent $+$ Librarian} \\
\cmidrule(lr){3-4} \cmidrule(lr){6-7}
Model & Role & Out (K) & E (kJ) & & Out (K) & E (kJ) \\
\midrule
\multirow{4}{*}{A3B} & planner & 9.20 & 5.03 & & 9.46 & 5.35 \\
 & navigator $\rightarrow$ librarian & 13.92 & 7.54 & & 6.02 & 3.57 \\
 & editor & 2.84 & 1.48 & & 2.81 & 1.46 \\
 & executor & 22.70 & 11.69 & & 24.12 & 12.46 \\
\midrule
\multirow{4}{*}{27B} & planner & 7.72 & 26.63 & & 7.49 & 26.40 \\
 & navigator $\rightarrow$ librarian & 10.50 & 36.63 & & 4.70 & 16.92 \\
 & editor & 2.55 & 8.82 & & 1.95 & 6.72 \\
 & executor & 14.66 & 49.82 & & 12.20 & 41.46 \\
\bottomrule
\end{tabular}
}

\caption{Per-role per-episode comparison of HyperAgent against its Librarian variant. We report the per-episode means of output tokens and energy at each role on the two backbones. The Librarian compresses the lookup role while leaving the other roles within $10\%$ on every metric.}
\label{tab:role-comparison-hyperagent}
\end{table}
\begin{table}[t!]
\centering
\setlength{\tabcolsep}{6pt}
\renewcommand{\arraystretch}{1.15}
\resizebox{\linewidth}{!}{
\begin{tabular}{@{}llrrcrr@{}}
\toprule
 & & \multicolumn{2}{c}{BOAD} & & \multicolumn{2}{c}{BOAD $+$ Librarian} \\
\cmidrule(lr){3-4} \cmidrule(lr){6-7}
Model & Role & Out (K) & E (kJ) & & Out (K) & E (kJ) \\
\midrule
\multirow{3}{*}{A3B} & orchestrator & 11.01 & 6.74 & & 8.79 & 5.23 \\
 & issue analyzer & 14.17 & 8.63 & & 14.31 & 8.62 \\
 & code\_navigator $\rightarrow$ librarian & 13.81 & 8.54 & & 1.93 & 1.08 \\
\midrule
\multirow{3}{*}{27B} & orchestrator & 6.66 & 24.23 & & 6.07 & 21.89 \\
 & issue analyzer & 12.11 & 44.48 & & 11.79 & 43.35 \\
 & code\_navigator $\rightarrow$ librarian & 5.63 & 20.79 & & 1.82 & 6.56 \\
\bottomrule
\end{tabular}
}

\caption{Per-role per-episode comparison of BOAD against its Librarian variant. We report the per-episode means of output tokens and energy at each role on the two backbones. The Librarian compresses the lookup role while leaving the other roles within $10\%$ on every metric.}
\label{tab:role-comparison-boad}
\end{table}

\subsection{\rev{Tightness of the duplicate-output bound}}
\label{app:dup-bound}

\rev{\S\ref{sec:redundancy-source} reports an upper bound on duplicate output rather than a point estimate, because a turn flagged as a duplicate read may also emit new reasoning that the accounting cannot separate out. The per-role numbers above indicate that the bound is tight, that is, that little of the flagged output carries content the episode still needs.}

\rev{The \Librarian removes only the duplicate lookups, yet the output of the role it replaces drops by 52 to 86\% across all eight comparisons (Table~\ref{tab:role-comparison-compact}), while Tables~\ref{tab:role-comparison-hyperagent} and~\ref{tab:role-comparison-boad} show the remaining roles stay within $10\%$ of the baseline on every metric. Had the flagged tokens mostly carried new task-relevant reasoning, some other role would have had to re-derive it once the duplicate lookups were gone, and none does. Pass rate is preserved as well (Table~\ref{tab:main-results}), so the removed output was not carrying content the episode depended on.}

\section{Pass Rate across Task Difficulty}
\label{app:difficulty-pass}

Table~\ref{tab:difficulty-pass} stratifies pass rate by the difficulty bins of \S\ref{sec:exp-main}, complementing the per-episode energy reported there. \rev{The \Librarian's pass rate is at or above the underlying MAS's in 12 of the 16 bins, and within $1$ point below it in three of the remaining four. The one larger deviation is the ${>}96$K bin of HyperAgent on 35B-A3B, $40.9$ against $22.7$, which is a difference of 4 tasks in a bin of 22. We treat these per-bin numbers as descriptive rather than as evidence of a statistically reliable difference, since the bins are small.}

\begin{table}[t!]
\centering
\resizebox{\linewidth}{!}{
\begin{tabular}{lrrrr}
\toprule
\multicolumn{1}{c}{} & \multicolumn{4}{c}{Reference max-input tokens (K)} \\
\cmidrule(lr){2-5}
\multirow{2}{*}{Method} & 0--32K & 32--64K & 64--96K & >96K \\
 & (n=36) & (n=312) & (n=130) & (n=22) \\
\midrule
\rowcolor{gray!12}
\multicolumn{5}{l}{\textbf{Qwen3.6-35B-A3B}} \\
HyperAgent & 83.3 & 73.4 & 47.7 & \textbf{40.9} \\
\hspace{1ex}$+$ Librarian & \textbf{86.1} & \textbf{74.7} & \textbf{51.5} & 22.7 \\
\hspace{1ex}$+$ LastNObservation & 80.6 & 67.3 & 32.3 & 13.6 \\
\hspace{1ex}$+$ Caveman & 77.8 & 73.1 & 50.8 & \textbf{40.9} \\
\hspace{1ex}$+$ Caveman $+$ Librarian & \textbf{86.1} & 73.1 & 50.0 & 31.8 \\
\midrule
BOAD & \textbf{88.9} & 76.6 & 53.8 & \textbf{36.4} \\
\hspace{1ex}$+$ Librarian & \textbf{88.9} & \textbf{78.8} & 55.4 & \textbf{36.4} \\
\hspace{1ex}$+$ LastNObservation & 83.3 & 71.5 & 45.4 & 18.2 \\
\hspace{1ex}$+$ Caveman & 86.1 & 73.7 & \textbf{57.7} & 27.3 \\
\hspace{1ex}$+$ Caveman $+$ Librarian & \textbf{88.9} & 77.9 & 53.8 & 31.8 \\
\midrule
\rowcolor{gray!12}
\multicolumn{5}{l}{\textbf{Qwen3.6-27B}} \\
HyperAgent & 83.3 & \textbf{75.6} & 57.7 & 36.4 \\
\hspace{1ex}$+$ Librarian & \textbf{86.1} & 75.0 & \textbf{65.4} & \textbf{45.5} \\
\hspace{1ex}$+$ LastNObservation & \textbf{86.1} & 75.0 & 42.3 & 18.2 \\
\hspace{1ex}$+$ Caveman & 83.3 & 73.1 & 56.2 & 40.9 \\
\hspace{1ex}$+$ Caveman $+$ Librarian & \textbf{86.1} & 72.4 & 54.6 & \textbf{45.5} \\
\midrule
BOAD & 86.1 & \textbf{77.6} & \textbf{60.8} & 31.8 \\
\hspace{1ex}$+$ Librarian & \textbf{88.9} & 76.6 & 60.0 & \textbf{45.5} \\
\hspace{1ex}$+$ LastNObservation & \textbf{88.9} & 77.2 & 55.4 & 27.3 \\
\hspace{1ex}$+$ Caveman & \textbf{88.9} & 75.3 & 57.7 & 40.9 \\
\hspace{1ex}$+$ Caveman $+$ Librarian & \textbf{88.9} & 75.6 & \textbf{60.8} & \textbf{45.5} \\
\bottomrule
\end{tabular}
}

\caption{Pass rate of the SWE agents with respect to the task difficulty. We report the pass rate of the SWE agents across difficulty classes. \rev{The Librarian's pass rate is at or above that of the underlying MAS in 12 of the 16 bins.}}
\label{tab:difficulty-pass}
\end{table}

\section{Librarian Invocation Frequency}
\label{app:invocation}

We examine how often the orchestrator invokes the \Librarian as the search problem grows. Stratifying episodes by the difficulty bins of \S\ref{sec:exp-main}, we count the per-episode \Librarian invocations. Figure~\ref{fig:librarian-wakes-by-bin} shows this mean count growing monotonically with difficulty across both backbones. We attribute the trend to the orchestrator scaling lookup frequency with the size of the search problem.

\begin{figure}[t!]
\centering
\input{figures/librarian_wakes_by_bin}
\caption{Mean Librarian invocations per episode across task difficulty. Each panel plots the per-episode mean Librarian invocation count for the Librarian-augmented MAS variants against the reference max-input-token bin. Bar color denotes the host family, and a hatch marks the Caveman variant. The invocation frequency grows monotonically with task difficulty in every setting.}
\label{fig:librarian-wakes-by-bin}
\end{figure}

\section{\rev{Paired Significance Tests}}
\label{app:significance}

\rev{Table~\ref{tab:significance} tests the \Librarian's two central claims on the full 500-task set of SWE-Bench Verified. Every comparison is paired task by task, and the two claims use different designs because they ask different questions: for cost we ask whether a reduction exists, and for pass rate we ask whether the underlying system's performance is preserved.}

\rev{\paragraph{Output tokens and energy.}
We test each cost reduction with a two-sided Wilcoxon signed-rank test~\citep{wilcoxon1945} on per-task paired differences. On both metrics the same 7 of 8 comparisons are significant and survive Holm correction~\citep{holm1979} for multiple comparisons; the single exception is HyperAgent$+$Caveman on 35B-A3B.}

\rev{\paragraph{Pass rate.}
For pass rate the claim is preservation, and a non-significant difference is not evidence of similarity, so we instead run one-sided non-inferiority tests for paired binary outcomes~\citep{liu2002noninferiority} at margins of 2 and 3.5 points. Following McNemar's design~\citep{mcnemar1947}, we count the tasks on which exactly one of the two systems passes. Non-inferiority holds in 5 of 8 comparisons at the 2-point margin and in all 8 at the 3.5-point margin.}

\begin{table*}[t!]
\centering
\revon
\setlength{\tabcolsep}{5pt}
\renewcommand{\arraystretch}{1.1}
\resizebox{\textwidth}{!}{
\begin{tabular}{l rr rr rrr}
\toprule
& \multicolumn{4}{c}{Cost reduction (Wilcoxon signed-rank)} & \multicolumn{3}{c}{Pass-rate non-inferiority (McNemar)} \\
\cmidrule(lr){2-5}\cmidrule(lr){6-8}
Comparison & $\Delta$out (K) & $p$ & $\Delta E$ (kJ) & $p$ & $\Delta$pass (pt) & $p_{NI}$ at 2\,pt & $p_{NI}$ at 3.5\,pt \\
\midrule
\rowcolor{gray!12}
\multicolumn{8}{l}{\textbf{Qwen3.6-35B-A3B}} \\
HyperAgent $\rightarrow$ $+$Lib & $+6.28$ & $<0.001$ & $+2.90$ & $<0.001$ & $+1.2$ & $0.035$ & $0.004$ \\
HyperAgent$+$Cav $\rightarrow$ $+$Cav$+$Lib & $+3.00$ & $0.067$ & $+1.36$ & $0.150$ & $+0.0$ & $0.123$ & $0.021$ \\
BOAD $\rightarrow$ $+$Lib & $+3.95$ & $<0.001$ & $+2.79$ & $<0.001$ & $+1.8$ & $0.005$ & $<0.001$ \\
BOAD$+$Cav $\rightarrow$ $+$Cav$+$Lib & $+4.10$ & $<0.001$ & $+2.43$ & $<0.001$ & $+2.0$ & $0.005$ & $<0.001$ \\
\midrule
\rowcolor{gray!12}
\multicolumn{8}{l}{\textbf{Qwen3.6-27B}} \\
HyperAgent $\rightarrow$ $+$Lib & $+9.06$ & $<0.001$ & $+30.38$ & $<0.001$ & $+2.2$ & $0.004$ & $<0.001$ \\
HyperAgent$+$Cav $\rightarrow$ $+$Cav$+$Lib & $+4.10$ & $<0.001$ & $+14.17$ & $<0.001$ & $-0.4$ & $0.183$ & $0.040$ \\
BOAD $\rightarrow$ $+$Lib & $+4.58$ & $<0.001$ & $+17.20$ & $<0.001$ & $+0.0$ & $0.087$ & $0.009$ \\
BOAD$+$Cav $\rightarrow$ $+$Cav$+$Lib & $+3.99$ & $<0.001$ & $+14.99$ & $<0.001$ & $+1.2$ & $0.015$ & $<0.001$ \\
\bottomrule
\end{tabular}
}

\caption{\rev{Paired significance tests on the full 500 tasks of SWE-Bench Verified. For cost, $\Delta$ is the baseline minus the \Librarian value, so positive values indicate a reduction, and $p$ comes from a two-sided Wilcoxon signed-rank test. For pass rate, $\Delta$pass is the \Librarian minus the baseline, and $p_{NI}$ is the one-sided p-value against the null hypothesis that the \Librarian is worse than the baseline by more than the stated margin, so a small $p_{NI}$ supports non-inferiority. All p-values below $0.001$ are reported as $<0.001$.}}
\label{tab:significance}
\end{table*}

\end{document}